\documentclass[12pt]{article}
\usepackage{bbm,authblk,indentfirst}
\usepackage{amsfonts}
\usepackage{amsmath}
\usepackage{amssymb}
\usepackage{amsthm}
\usepackage{cite}
\usepackage{mathrsfs}
\usepackage[colorlinks]{hyperref}
\usepackage{graphicx}
\usepackage[left=2.5cm,right=2.5cm,bottom=2.8cm,top=2.8cm]{geometry}

\theoremstyle{plain}
\newtheorem{theorem}{Theorem}[section]
\newtheorem{corollary}[theorem]{Corollary}
\newtheorem{lemma}[theorem]{Lemma}
\newtheorem{assume}[theorem]{Assumption}
\newtheorem{prop}[theorem]{Proposition}
 \theoremstyle{remark}
\newtheorem{remark}[theorem]{Remark}
 \theoremstyle{definition}
 \newtheorem{definition}[theorem]{Definition}
\newtheorem{problem}[theorem]{Riemann--Hilbert Problem}
\numberwithin{equation}{section}
\allowdisplaybreaks[2]
\begin{document}
\title{\LARGE\bf Real-Valued Vector Modified Korteweg--de Vries Equation: Solitons Featuring Multiple Poles}

\renewcommand*{\Affilfont}{\small\it}
\renewcommand{\Authands}{, }
\author[1]{Zhenzhen Yang}
\author[1]{Huan Liu\thanks{ E-mail: liuhuan@zzu.edu.cn}}
\author[2]{Jing Shen}
\affil[1]{School of Mathematics and Statistics, Zhengzhou University, Zhengzhou, Henan 450001, People's  Republic  of China}
\affil[2]{School of Mathematics and Statistics,  Henan  University  of  Technology,  Zhengzhou, Henan  450001,  People's  Republic  of China}
\date{}

\maketitle
\begin{abstract}
\vspace{0.5cm}
We delve into the inverse scattering transform of the real-valued vector modified Korteweg--de Vries equation, emphasizing the challenges posed by $N$ pairs of higher-order poles in the transmission coefficient and the enhanced spectral symmetry stemming from real-valued constraints. Utilizing the generalized vector cross product, we formulate an $(n+1)\times(n+1)$ matrix-valued Riemann--Hilbert problem to tackle the complexities inherent in multi-component systems. We subsequently demonstrate the existence and uniqueness of solutions for a singularity-free equivalent problem, adeptly handling the intricacies of multiple poles. In reflectionless cases, we reconstruct multi-pole soliton solutions through a system of linear algebraic equations.

\end{abstract}
\newpage
\section{Introduction}\setcounter{equation}{0}

Solitons, renowned for their stability in wave propagation, play a crucial role in nonlinear physics. They are vital in various fields such as fiber-optic communications\cite{Newell1985,Hasegawa1995}, plasma physics\cite{Kuznetsov1986,Pathak2017,Cheemaa2020}, Bose--Einstein condensation\cite{Pitaevskii2003} and  oceanography\cite{Osborne2010}. Solitons are solutions to nonlinear wave equations where the balance between dispersion and nonlinearity is achieved, a characteristic typical of integrable systems. To find soliton solutions, various methods are employed, including the inverse scattering transform (IST)\cite{Gardner1967,Ablowitz1974,Ablowitz1981}, the Hirota bilinear method\cite{Hirota1971,Hirota2004}, the Darboux transformation\cite{Matveev1991,Gu2005,Guo2013},  and the Riemann--Hilbert (RH) method\cite{Novikov1984,Its2003}.

In the context of IST theory, the poles of the transmission coefficient are essential for generating soliton solutions in integrable equations. The presence of $N$ pairs of simple conjugate poles directly leads to the formation of $N$-soliton solutions. This principle extends to the study of reflectionless solutions, characterized by $N$ pairs of higher-order poles, known as multi-pole solutions. There has been extensive research on integrable equations associated with the Ablowitz--Kaup--Newell--Segur (AKNS) spectral problem, encompassing the nonlinear Schr\"odinger  equation \cite{Zakharov1972,Olmedilla1987,Aktosun2007,Schiebold2017,Zhang2020a,Bilman2019,Bilman2021}, the modified Korteweg--de Vries equation \cite{Wadati1982,Zhang2020b}, and the sine-Gordon equation \cite{Tsuru1984}. Furthermore, studies have also delved into discrete integrable equations, such as the discrete sine-Gordon equation \cite{Fan2022} and the Ablowitz--Ladik equation\cite{Liu2024a}, as well as integrable equations  associated with third and fourth order matrix spectral problems, including the Sasa--Satsuma equation\cite{Liu2024b} and the spin-1 Gross--Pitaevskii equation\cite{Liu2024c}.

The complexity and dynamics of integrable systems are enhanced by multi-component coupling, which leads to interactions between components and results in phenomena not present in single-component systems. Investigating these couplings reveals new algebraic structures that are fundamental to the advancement of integrable system theory. This exploration not only deepens our understanding of solitons but also enriches the mathematical and physical frameworks that describe them. Consequently, many researchers have investigated multi-component integrable systems\cite{Yao2004,Zhang2008,Geng2014,Tian2017,Chang2018,Pelinovsky2018,Geng2019,Wurile2019,Adamopoulou2020,Xiao2021,Malham2022,Liu2023a,Ye2023,Wong2024}, which is both essential and significant.
However, the literature on higher-order poles solutions to integrable equations associated with  multi-component AKNS spectral problems remains scarce.

This paper focuses on the vector modified Korteweg--de Vries(vmKdV) equation\cite{Liu2016}
\begin{equation}\label{eq:vmKdVe}
\mathbf{q}_{t}+\mathbf{q}_{xxx}+3\left(\mathbf{q}_{x}\mathbf{q}^{\mathrm{T}}\mathbf{q}+\mathbf{q}\mathbf{q}^{\mathrm{T}}\mathbf{q}_{x}\right)=\mathbf{0},
\end{equation}
where $\mathbf{q}=(\mathrm{q}_{1},\ldots,\mathrm{q}_{n})^{\mathrm{T}}$ is an $n$ dimensional real-valued vector  function of independent variables $x$ and $t$.
Wang and Han \cite{Wang2020} derived soliton solutions to the vmKdV equation \eqref{eq:vmKdVe} by dressing the RH problem associated with single-order zeros. Our work, however, aims to establish an inverse scattering analysis based on an $(n+1)\times(n+1)$  matrix RH problem that includes several residue conditions at $N$ pairs of multiple poles.  We introduce a generalized cross product operator in $\mathbb{C}^{n+1}$, thereby laying the groundwork for the subsequent analysis of the discrete spectrum.  We also rigorously demonstrate the existence and uniqueness of the solution to this RH problem.  To our knowledge, no previous studies have investigated multiple higher-order pole solutions of the vmKdV equation \eqref{eq:vmKdVe} within the framework of the RH problem associated  with $(n+1)\times(n+1)$ matrix spectral problem.

The research process  is as follows: In Section \ref{sec:dir}, we investigate the direct scattering problem, constructing a mapping from the initial data to the scattering data and analyzing the discrete spectrum related to the $N$ pairs of multiple zeros. In Section \ref{sec:inv}, we explore the inverse scattering problem, constructing a mapping from the scattering data to  an $(n+1)\times(n+1)$ matrix RH problem in the absence of $N$ pairs of multiple poles. In the reflectionless case, we construct a concise linear algebraic system, allowing us to obtain  multi-pole solutions solving this linear system, with more explicit solutions derived by appropriately selecting parameters.

Throughout this paper, we adhere to a set of defined notations to maintain clarity and consistency. The complex conjugate of a complex number $\lambda$ is denoted by $\bar{\lambda}$. For a complex-valued matrix $\mathbf{A}$, $\mathrm{ad}[\mathbf{A}]$  signifies the adjugate of $\mathbf{A}$, $\mathbf{\bar{A}}$ denotes the element-wise complex conjugate, $\mathbf{A}^\mathrm{T}$  indicates  the transpose, and $\mathbf{A}^{\dag}$ represents the conjugate transpose. The commutator of two $(n+1)\times(n+1)$ matrices $\mathbf{A}$ and $\mathbf{B}$ is defined as $[\mathbf{A},\mathbf{B}]=\mathbf{A}\mathbf{B}-\mathbf{B}\mathbf{A}$. An $(n+1)\times(n+1)$ matrix $\mathbf{A}$ is represented in block form as follows:
 \begin{equation*}
 \begin{aligned}
 \mathbf{A}=&\begin{pmatrix}
\mathbf{A}_{11}&\mathbf{A}_{12}&\cdots&\mathbf{A}_{1(n+1)}\\
\mathbf{A}_{21}&\mathbf{A}_{22}&\cdots&\mathbf{A}_{2(n+1)}\\
\vdots&\vdots&\cdots&\vdots\\
\mathbf{A}_{(n+1)1}&\mathbf{A}_{(n+1)2}&\cdots&\mathbf{A}_{(n+1)(n+1)}
\end{pmatrix}\\
=&(\mathbf{A}_{1},\mathbf{A}_2,
\ldots,\mathbf{A}_{n+1})
=(\mathbf{A}_{\mathrm{L}},\mathbf{A}_{\mathrm{R}})
=\begin{pmatrix}
\mathbf{A}_{\mathrm{UL}}&\mathbf{A}_{\mathrm{UR}}\\
\mathbf{A}_{\mathrm{DL}}&\mathbf{A}_{\mathrm{DR}}
\end{pmatrix},
\end{aligned}
\end{equation*}
where $\mathbf{A}_{ij}$ refers to the $(i,j)$-entry of the matrix, $\mathbf{A}_{j}$ represents the $j$-th column, $\mathbf{A}_{\mathrm{L}}$ refers to the first column, $\mathbf{A}_{\mathrm{R}}$ represents the last $n$ columns, $\mathbf{A}_{\mathrm{UL}}$ is a scalar, and $\mathbf{A}_{\mathrm{DR}}$ is an $n\times n$ matrix. The identity matrix of appropriate size is denoted by  $\mathbf{I}$, and $\mathbb{C^{\pm}}$ represents the complex plane divided into upper and lower half-planes. For a vector-valued function $\mathbf{f}(x,t;\lambda)$, we define $\mathbf{f}^{(h)}(x,t;\lambda)=\partial_{\lambda}^{h}\mathbf{f}(x,t;\lambda)$, $\mathbf{f}^{(h)}(x,t;\lambda_{0})=\partial_{\lambda}^{h}\mathbf{f}(x,t;\lambda)|_{\lambda=\lambda_{0}}$. For simplicity, we occasionally omit the variables  $x$ and $t$.
\section{Direct scattering problem}\label{sec:dir}
\subsection{Jost solutions}
The vmKdV equation \eqref{eq:vmKdVe}  is associated with the Lax pair:
\begin{subequations}\label{eq:Laxpair}
\begin{align}
&\psi_{x}=(-\mathrm{i}\lambda\sigma+\mathbf{Q})\psi,\\
&\psi_{t}=(-4\mathrm{i}\lambda^{3}\sigma+\tilde{\mathbf{Q}})\psi,
\end{align}
\end{subequations}
where
 \begin{equation}\label{uvdef}
 \begin{aligned}
 &\sigma=\begin{pmatrix}
1&\mathbf{0}\\
\mathbf{0}&-\mathbf{I}_{n\times n}
\end{pmatrix},\quad
\mathbf{Q}=\begin{pmatrix}
0&-\mathbf{q}^\mathrm{T}\\
\mathbf{q}&\mathbf{0}_{n\times n}
\end{pmatrix},\\
&\tilde{\mathbf{Q}}=4\lambda^{2} \mathbf{Q}+2\mathrm{i}\lambda\sigma\left(\mathbf{Q}_{x}-\mathbf{Q}^{2}\right)+2\mathbf{Q}^{3}-\mathbf{Q}_{xx}+\left[\mathbf{Q}_{x},\mathbf{Q}\right],
\end{aligned}
\end{equation}
and $\psi(x,t;\lambda)$ is an $(n+1)\times(n+1)$ matrix-valued function depending on $x, t$, and the spectral parameter $\lambda \in \mathbb{C}$. The vmKdV equation \eqref{eq:vmKdVe} is equivalent to the zero-curvature condition:
\begin{equation}\label{eq:zcq}
\mathbf{Q}_{t}-\tilde{\mathbf{Q}}_{x}+[-\mathrm{i}\lambda\sigma+\mathbf{Q},-4\mathrm{i}\lambda^{3}\sigma+\tilde{\mathbf{Q}}]=\mathbf{0}.
\end{equation}

We are in search of a solution $\mathbf{q}(x,t)$ to Eq.\eqref{eq:vmKdVe} that decays rapidly to zero as $x$ becomes sufficiently large for any $(x,t) \in \mathbb{R} \times \mathbb{R}^{+}$. Subsequently, we examine Eq.\eqref{eq:Laxpair} and seek two fundamental solution matrices $\psi_{\pm}(x,t;\lambda)$, which satisfy the boundary conditions for $\lambda \in \mathbb{R}$:
\begin{equation}\label{eq:Psiasy}
\psi_{\pm}(x,t;\lambda) = \mathrm{e}^{-\mathrm{i}\theta(x,t;\lambda)\sigma} + o(1), \quad x \rightarrow \pm\infty,
\end{equation}
where $\theta(x,t;\lambda) = \lambda x + 4\lambda^{3}t$. In this context, $\psi_{\pm}(x,t;\lambda)$ are known as the Jost solutions of Eq.\eqref{eq:Laxpair}.

Define
\begin{equation}
\label{mJost}
\mu_{\pm}(x,t;\lambda) = \psi_{\pm}(x,t;\lambda) \mathrm{e}^{\mathrm{i}\theta(x,t;\lambda)\sigma},
\end{equation}
then we have the following system of equations:
\begin{equation}\label{eq:muLaxpair}
\begin{aligned}
&\partial_{x}\mu_{\pm}(x,t;\lambda) = [\mu_{\pm}(x,t;\lambda), \mathrm{i}\lambda\sigma] + \mathbf{Q}(x,t)\mu_{\pm}(x,t;\lambda), \\
&\partial_{t}\mu_{\pm}(x,t;\lambda) = [\mu_{\pm}(x,t;\lambda), 4\mathrm{i}\lambda^{3}\sigma] + \tilde{\mathbf{Q}}(x,t;\lambda)\mu_{\pm}(x,t;\lambda), \\
&\lim_{x\rightarrow\pm\infty} \mu_{\pm}(x,t;\lambda) = \mathbf{I}.
\end{aligned}
\end{equation}
The functions $\mu_{\pm}(x,t;\lambda)$ can be expressed in the form of Volterra integral equation:
\begin{equation}\label{eq:mujifen}
\mu_{\pm}(x,t;\lambda) = \mathbf{I} + \int_{\pm\infty}^{x} \mathrm{e}^{\mathrm{i}\lambda(\xi-x)\hat{\sigma}}[\mathbf{Q}(\xi,t)\mu_{\pm}(\xi,t;\lambda)]\,\mathrm{d}\xi,
\end{equation}
where $\hat{\sigma} \mathbf{X} = [\sigma, \mathbf{X}]$ and $\mathrm{e}^{\hat{\sigma}}\mathbf{X} = \mathrm{e}^{\sigma}\mathbf{X}\mathrm{e}^{-\sigma}$.

Drawing on the theory of Volterra integral equations and the boundedness of the integral factor, we can establish the following theorem:
\begin{theorem} Suppose that $\mathbf{q}(\cdot, t) \in \mathrm{L}^{1}(\mathbb{R})$ for a fixed $t$. Then, the modified eigenfunctions $\mu_{\pm}(x,t;\lambda)$ defined in Eq.\eqref{eq:mujifen} are well-defined for $\lambda \in \mathbb{R}$. Specifically, $\mu_{\mathrm{-L}}(x,t;\lambda)$ and $\mu_{\mathrm{+R}}(x,t;\lambda)$ can be analytically continued to the upper half-plane $\mathbb{C}^{+}$, while $\mu_{\mathrm{+L}}(x,t;\lambda)$ and $\mu_{\mathrm{-R}}(x,t;\lambda)$ can be analytically continued to the lower half-plane $\mathbb{C}^{-}$. Within the interior of their respective domains of analyticity, $\mu_{\pm}(x,t;\lambda)$ remain bounded for $x \in \mathbb{R}$. It is noteworthy that the functions $\psi_{\pm}(x,t;\lambda)$ exhibit analogous properties of analyticity.
\end{theorem}
\subsection{Symmetry and asymptotic behavior}
The spectral problem \eqref{eq:Laxpair} is characterized by a pair of symmetries that play a crucial role in shaping its solutions. These symmetries are encapsulated by the transformations $\lambda \rightarrow \bar{\lambda}$, which corresponds to complex conjugation, and $\lambda \rightarrow -\lambda$, representing reflection about the origin in the complex plane. Building on these symmetries, and particularly leveraging the skew-symmetry property of the matrix $\mathbf{Q}$, which is expressed as $\mathbf{Q}^\mathrm{T} = -\mathbf{Q}$, we derive significant implications for the fundamental matrix solutions $\psi_\pm(x,t;\lambda)$ of the Lax pair \eqref{eq:Laxpair}.

\begin{prop}The fundamental matrix solutions $\psi_\pm(x,t;\lambda)$ of the Lax pair \eqref{eq:Laxpair} display the symmetries:
\begin{equation}
\label{eq:Jostsym}
\psi_{\pm}^{-1}(x,t;\lambda)=\psi_{\pm}^{\dag}(x,t;\bar{\lambda})=\psi_{\pm}^{\mathrm{T}}(x,t;-\lambda),\quad \lambda\in\mathbb{R}.
\end{equation}
Furthermore,
\begin{equation}
\label{eq:mJostsym}
\mu_{\pm}^{-1}(x,t;\lambda)=\mu_{\pm}^{\dag}(x,t;\bar{\lambda})=\mu_{\pm}^{\mathrm{T}}(x,t;-\lambda),\quad \lambda\in\mathbb{R}.
\end{equation}
\end{prop}
Given that $\mathbf{Q}$ is traceless, Abel's Theorem implies that $\partial_{x}\mathrm{det}[\mu_{\pm}(x,t;\lambda)] = 0$. By combining Eq.\eqref{mJost} with Eq.\eqref{eq:muLaxpair}, we deduce that
\begin{equation}
\label{eq:mudet}
\mathrm{det}[\mu_{\pm}(x,t;\lambda)]=1,\quad   \mathrm{det}[\psi_{\pm}(x,t;\lambda)]=\mathrm{e}^{(n-1)\mathrm{i}\theta(x,t;\lambda)},\quad\lambda\in\mathbb{R}.
\end{equation}
Since both $\psi_+(x,t;\lambda)$ and $\psi_-(x,t;\lambda)$ are fundamental solutions of the Lax pair \eqref{eq:Laxpair}, an $n \times n$ scattering matrix $\mathbf{S}(\lambda)$ exists, which is independent of $x$ and $t$, and satisfies the relationship:
\begin{equation}
\label{eq:Jostjump}
\psi_-(x,t;\lambda)=\psi_+(x,t;\lambda) \mathbf{S}(\lambda),\quad \lambda\in\mathbb{R}.
\end{equation}
Alternatively,
\begin{equation}
\label{eq:mJostjump}
\mu_{-}(x,t;\lambda)=\mu_{+}(x,t;\lambda)\mathrm{e}^{-\mathrm{i}\theta(x,t;\lambda)\hat{\sigma}}\mathbf{S}(\lambda),  \quad\lambda\in\mathbb{R}.
\end{equation}
In light of Eq.\eqref{eq:Jostsym} and Eq.\eqref{eq:Jostjump}, we find that $\mathbf{S}(\lambda)$ satisfies the following properties:
\begin{equation}
\label{eq:ssym}
\mathrm{det}[\mathbf{S}(\lambda)]=1,\quad\mathbf{S}^{-1}(\lambda)=\mathbf{S}^{\dag}(\bar{\lambda})=\mathbf{S}^{\mathrm{T}}(-\lambda),\quad\lambda\in\mathbb{R}.
\end{equation}
Consequently, the relationships between the components of $\mathbf{S}(\lambda)$ are derived as:
\begin{equation}
\mathbf{S}_{\mathrm{UL}}^{\dag}(\bar{\lambda})=\mathrm{det}[\mathbf{S}_{\mathrm{DR}}(\lambda)],\quad\mathbf{S}_{\mathrm{DL}}^{\dag}(\bar{\lambda})=-\mathbf{S}_{\mathrm{UR}}(\lambda)\mathrm{ad}[\mathbf{S}_{\mathrm{DR}}(\lambda)].
\end{equation}
Based on the above relationships, $\mathbf{S}(\lambda)$ can be expressed in the form:
\begin{equation}
\label{eq:S}
\mathbf{S}(\lambda)=\begin{pmatrix}
\mathrm{det}[\mathbf{a}^{\dag}(\bar{\lambda})]&\mathbf{b}(\lambda)\\
-\mathrm{ad}[\mathbf{a}^{\dag}(\bar{\lambda})]\mathbf{b}^{\dag}(\bar{\lambda})&\mathbf{a}(\lambda)\\
\end{pmatrix},
\quad\lambda\in\mathbb{R}.
\end{equation}
In addition,
\begin{equation}
\label{eq:absym}
\mathbf{a}^{\dag}(\bar{\lambda})=\mathbf{a}^{\mathrm{T}}(-\lambda),\quad
\mathbf{b}^{\dag}(\bar{\lambda})=\mathbf{b}^{\mathrm{T}}(-\lambda),\quad
\quad\lambda\in\mathbb{R}.
\end{equation}
According to Eqs.\eqref{eq:mujifen} and \eqref{eq:mJostjump}, the $\mathbf{a}(\lambda)$ and $\mathbf{b}(\lambda)$ in the scattering matrix $\mathbf{S}(\lambda)$  can be expressed in integral form as:
\begin{equation}
\begin{aligned}
    &\mathbf{a}(\lambda)=\mathbf{I}+\int^{+\infty}_{-\infty}\mathbf{q}(x,0)\mu_{\mathrm{-UR}}(x,0;\lambda)\mathrm{d}x,\\
    &\mathbf{b}(\lambda)=\int^{+\infty}_{-\infty}-\mathrm{e}^{2\mathrm{i}\lambda x}\mathbf{q}^{\mathrm{T}}(x,0)\mu_{\mathrm{-DR}}(x,0;\lambda)\mathrm{d}x.
\end{aligned}
\end{equation}
Given that $\mathbf{q}(x,0)\in\mathrm{L}^{1}(\mathbb{R})$, $\mathbf{a}(\lambda)$ and $\mathbf{b}(\lambda)$ are well-defined for $\lambda\in\mathbb{R}$, and $\mathbf{a}(\lambda)$ can be analytically continued onto $\mathbb{C}^{-}$. Furthermore, based on Eqs.\eqref{mJost}, \eqref{eq:mJostsym}, \eqref{eq:mJostjump} and \eqref{eq:S}, it can be concluded that $\mathbf{a}(\lambda)$ and $\mathbf{b}(\lambda)$ can be expressed in terms of $\mu_{\pm}(x,t;\lambda)$ or $\psi_{\pm}(x,t;\lambda)$. Indeed,
\begin{subequations}
\label{eq:ab1}
\begin{alignat}{2}
&\mathbf{a}(\lambda)=(\mu_{\mathrm{+R}})^{\dag}(\bar{\lambda})\mu_{\mathrm{-R}}(\lambda)=(\psi_{\mathrm{+R}})^{\dag}(\bar{\lambda})\psi_{\mathrm{-R}}(\lambda),&\lambda\in\mathbb{C^{-}}\cup\mathbb{R},\\
&\mathrm{det}[\mathbf{a}(\lambda)]=\mathrm{det}[\mu_{\mathrm{+L}}(\lambda),\mu_{\mathrm{-R}}(\lambda)]=\mathrm{e}^{-(n-1)\mathrm{i}\theta(\lambda)}\mathrm{det}[\psi_{\mathrm{+L}}(\lambda),\psi_{\mathrm{-R}}(\lambda)],&\lambda\in\mathbb{C^{-}}\cup\mathbb{R},\\
&\mathbf{a}^{\dag}(\bar{\lambda})=(\mu_{\mathrm{-R}})^{\dag}(\bar{\lambda})\mu_{\mathrm{+R}}(\lambda)=(\psi_{\mathrm{-R}})^{\dag}(\bar{\lambda})\psi_{\mathrm{+R}}(\lambda),&\lambda\in\mathbb{C^{+}}\cup\mathbb{R},\\
&\mathrm{det}[\mathbf{a}^{\dag}(\bar{\lambda})]=\mathrm{det}[\mu_{\mathrm{-L}}(\lambda),\mu_{\mathrm{+R}}(\lambda)]=\mathrm{e}^{-(n-1)\mathrm{i}\theta(\lambda)}\mathrm{det}[\psi_{\mathrm{-L}}(\lambda),\psi_{\mathrm{+R}}(\lambda)],&\lambda\in\mathbb{C^{+}}\cup\mathbb{R},\\
&\mathbf{b}(\lambda)=\mathrm{e}^{2\mathrm{i}\theta(\lambda)}(\mu_{\mathrm{+L}})^{\dag}(\bar{\lambda})\mu_{\mathrm{-R}}=(\psi_{\mathrm{+L}})^{\dag}(\bar{\lambda})\psi_{\mathrm{-R}}(\lambda),&\lambda\in\mathbb{R},\\
&-\mathrm{ad}[\mathbf{a}^{\dag}(\bar{\lambda})]\mathbf{b}^{\dag}(\bar{\lambda})=\mathrm{e}^{-2\mathrm{i}\theta(\lambda)}(\mu_{\mathrm{+R}})^{\dag}(\bar{\lambda})\mu_{\mathrm{-L}}(\lambda)=(\psi_{\mathrm{+R}})^{\dag}(\bar{\lambda})\psi_{\mathrm{-L}}(\lambda),&\lambda\in\mathbb{R}.
\end{alignat}
\end{subequations}

By substituting the Wentzel--Kramers--Brillouin expansion of $\mu_{\pm}(x,t;\lambda)$ into Eq.\eqref{eq:muLaxpair} and systematically collecting the terms of order  $\lambda^{j}$, we arrive at the following asymptotic results:
\begin{theorem}As $\lambda\rightarrow\infty$ within the relevant analytic region of $\mu_{\pm}(x,t;\lambda)$,
\begin{equation}
\begin{aligned}
&(\mu_{\mathrm{+L}}(x,t;\lambda),\mu_{\mathrm{-R}}(x,t;\lambda))=\mathbf{I}+\mathrm{O}(\lambda^{-1}),\quad&\lambda\in\mathbb{C^{-}}\rightarrow\infty,\\
&(\mu_{\mathrm{-L}}(x,t;\lambda),\mu_{\mathrm{+R}}(x,t;\lambda))=\mathbf{I}+\mathrm{O}(\lambda^{-1}),\quad&\lambda\in\mathbb{C^{+}}\rightarrow\infty.
\end{aligned}
\end{equation}
Consequently,
\begin{equation}
\label{eq:ab2}
\begin{aligned}
&\mathbf{a}(\lambda)=\mathbf{I}+\mathrm{O}(\lambda^{-1}),\quad&\lambda\rightarrow\infty,\\
&\mathbf{b}(\lambda)=\mathrm{O}(\lambda^{-1}),\quad&\lambda\rightarrow\infty.
\end{aligned}
\end{equation}
\end{theorem}

We define the reflection coefficient as
\begin{equation}
\label{eq:fanshe}
\mathbf{\gamma}(\lambda)=\mathbf{b}(\lambda)\mathbf{a}^{-1}(\lambda),\quad\lambda\in\mathbb{R}.
\end{equation}
The transmission coefficient is given by $\frac{1}{\mathrm{det}[\mathbf{a}(\lambda)]}$, and it follows that $\mathbf{\gamma}^{\dag}(\bar{\lambda})=\mathbf{\gamma}^{\mathrm{T}}(-\lambda)$. As $\lambda\rightarrow\infty$, we find that $\mathbf{\gamma}(\lambda)=\mathrm{O}(\lambda^{-1})$.

To further elucidate the symmetries inherent in  $\mu_
{\pm}(x,t;\lambda)$, we now introduce the definitions of two pivotal operators. These operators are essential for deciphering the algebraic frameworks that underpin our forthcoming analyses.
\begin{definition}(Generalized Cross Product) For all $\mathbf{u}_{1},\ldots,\mathbf{u}_{n}\in \mathbb{C}^{n+1}$, define
\begin{equation}
\label{eq:Gdef}
\mathcal{G}[\mathbf{u}_{1},\ldots,\mathbf{u}_{n}]=\sum_{j=1}^{n+1}\mathrm{det}(\mathbf{u}_{1},\ldots,\mathbf{u}_{n},\mathbf{e}_{j})\mathbf{e}_{j},
\end{equation}
where ${\mathbf{e}_{1},\ldots,\mathbf{e}_{n+1}}$ are the standard basis vectors of $\mathbb{R}^{n+1}$.
\end{definition}
\begin{definition}For all $\mathbf{u}_{1},\ldots,\mathbf{u}_{n+1}\in \mathbb{C}^{n+1}$, define
\begin{equation}
\label{eq:Gu}
\mathscr{G}[\mathbf{u}_{1},\ldots,\mathbf{u}_{n+1}]=-\sum_{l=1}^{n+1}\sum_{j=1}^{n+1}\mathrm{det}
\left(
\begin{array}{cc}
\mathbf{u}&\mathbf{e}_{j}\\
\mathbf{e}_{l}^{\mathrm{T}}&0
\end{array}
\right)
\mathbf{e}_{j}\mathbf{e}_{l}^{\mathrm{T}},
\end{equation}
where $\mathbf{u}=(\mathbf{u}_{1},\ldots,\mathbf{u}_{n+1})$. Thus, for $l=1,\ldots,n+1$, the $l$-th column of $\mathscr{G}[\mathbf{u}_{1},\ldots,\mathbf{u}_{n+1}]$ is given by
\begin{equation}
\label{eq:Gldef}
\mathscr{G}_{l}[\mathbf{u}_{1},\ldots,\mathbf{u}_{n+1}]=\mathscr{G}[\mathbf{u}_{1},\ldots,\mathbf{u}_{n+1}]\mathbf{e}_{l}=-\sum_{j=1}^{n+1}\mathrm{det}
\left(
\begin{array}{cc}
\mathbf{u}&\mathbf{e}_{j}\\
\mathbf{e}_{l}^{\mathrm{T}}&0
\end{array}
\right)
\mathbf{e}_{j}.
\end{equation}
\end{definition}
 By direct calculations, it is straightforward to verify the following relationship among the adjugate matrix $\mathrm{ad}[\cdot]$, the generalized cross product $\mathcal{G}[\cdot]$ and the operator $\mathscr{G}[\cdot]$:
\begin{equation}
\label{eq:adu}
\mathrm{ad}[\mathbf{u}]=
\left(
\begin{array}{cccc}
(-1)^{n}\mathcal{G}^{\mathrm{T}}[\mathbf{u}_{2},\ldots,\mathbf{u}_{n+1}]\\
(-1)^{n-1}\mathcal{G}^{\mathrm{T}}[\mathbf{u}_{1},\mathbf{u}_{3},\ldots,\mathbf{u}_{n+1}]\\
\quad\vdots\\
(-1)^{n+1-j}\mathcal{G}^{\mathrm{T}}[\mathbf{u}_{1},\ldots,\mathbf{u}_{j-1},\mathbf{u}_{j+1},\ldots,\mathbf{u}_{n+1}]\\
\quad\vdots\\
\mathcal{G}^{\mathrm{T}}[\mathbf{u}_{1},\ldots,\mathbf{u}_{n}]
\end{array}
\right)
=\mathscr{G}^{\mathrm{T}}[\mathbf{u}_{1},\ldots,\mathbf{u}_{n+1}].
\end{equation}
\begin{lemma}\label{lemu}
For all $\mathbf{u}_{1},\ldots,\mathbf{u}_{n}, \mathbf{v}_{1},\ldots,\mathbf{v}_{n}\in \mathbb{C}^{n+1}$,
\begin{equation}
(\mathscr{G}_{1},\ldots,\mathscr{G}_{n})[\mathbf{u}_{1},\ldots,\mathbf{u}_{n},\mathcal{G}[\mathbf{v}_{1},\ldots,\mathbf{v}_{n}]]=\mathbf{v}\mathrm{ad}[\mathbf{u}_{(1)}^{\mathrm{T}}\mathbf{v}],
\end{equation}
where $\mathbf{u}_{(1)}=(\mathbf{u}_{1},\ldots,\mathbf{u}_{n})$ and $\mathbf{v}=(\mathbf{v}_{1},\ldots,\mathbf{v}_{n})$.(see Lemma 3.3 in Ref.\cite{Liu2023b})
\end{lemma}
According to Eq.\eqref{eq:adu},  it is evident that
\begin{equation}
\begin{aligned}
\label{eq:admu}
\mathrm{ad}[\mu_{\pm}(\lambda)]=&
\left(
\begin{array}{cccc}
(-1)^{n}\mathcal{G}^{\mathrm{T}}[\mu_{\pm2}(\lambda),\ldots,\mu_{\pm (n+1)}(\lambda)]\\
(-1)^{n-1}\mathcal{G}^{\mathrm{T}}[\mu_{\pm1}(\lambda),\mu_{\pm3}(\lambda),\ldots,\mu_{\pm (n+1)}(\lambda)]\\
\quad\vdots\\
(-1)^{n+1-j}\mathcal{G}^{\mathrm{T}}[\mu_{\pm1}(\lambda),\ldots,\mu_{\pm (j-1)}(\lambda),\mu_{\pm (j+1)}(\lambda),\ldots,\mu_{\pm (n+1)}(\lambda)]\\
\quad\vdots\\
\mathcal{G}^{\mathrm{T}}[\mu_{\pm1}(\lambda),\ldots,\mu_{\pm n}(\lambda)]
\end{array}
\right)\\
=&\mathscr{G}^{\mathrm{T}}[\mu_{\pm1}(\lambda),\ldots,\mu_{\pm (n+1)}(\lambda)].
\end{aligned}
\end{equation}
When $\lambda\in\mathbb{R}$, $\det[\mu_\pm(x,t;\lambda)]=1$.  Based on Eq.\eqref{eq:mJostsym}, the following expression can be derived:
\begin{equation}
\label{eq:sym1}
[\mathrm{ad}[\mu_{\pm}(x,t;\lambda)]]^{\mathrm{T}}=\bar{\mu}_{\pm}(x,t;\bar{\lambda}).
\end{equation}
Since $[[\mathrm{ad}[\mu_{+}(x,t;\lambda)]]^{\mathrm{T}}]_{\mathrm{L}}$ and $\bar{\mu}_{\mathrm{+L}}(x,t;\bar{\lambda})$ are analytic in $\mathbb{C}^{+}$, the equality $[[\mathrm{ad}[\mu_{+}(x,t;\lambda)]]^{\mathrm{T}}]_{\mathrm{L}}=\bar{\mu}_{\mathrm{+L}}(x,t;\bar{\lambda})$ can be analytically continued to the domain $\mathbb{C}^{+}$. Similarly, since $[[\mathrm{ad}[\mu_{-}(x,t;\lambda)]]^{\mathrm{T}}]_{\mathrm{L}}$ and $\bar{\mu}_{\mathrm{-L}}(x,t;\bar{\lambda})$ are analytic in $\mathbb{C}^{-}$, the equality $[[\mathrm{ad}[\mu_{-}(x,t;\lambda)]]^{\mathrm{T}}]_{\mathrm{L}}=\bar{\mu}_{\mathrm{-L}}(x,t;\bar{\lambda})$ can be analytically continued to the domain $\mathbb{C}^{-}$.

Therefore, based on Eqs.\eqref{eq:Gldef}, \eqref{eq:admu} and \eqref{eq:sym1}, we can deduce that
\begin{equation}
\begin{aligned}
\label{eq:sym2}
&\bar{\mu}_{\mathrm{+L}}(x,t;\bar{\lambda})=[[\mathrm{ad}[\mu_{+}(x,t;\lambda)]]^{\mathrm{T}}]_{\mathrm{L}}\\
=&[\mathscr{G}[\mu_{+1}(x,t;\lambda),\ldots,\mu_{+(n+1)}(x,t;\lambda)]]_{\mathrm{L}}\\
=&\mathscr{G}_{1}[\mu_{+1}(x,t;\lambda),\ldots,\mu_{+(n+1)}(x,t;\lambda)]\\
=&(-1)^{n}\sum_{j=1}^{n+1}\mathrm{det}(\mu_{\mathrm{+R}}(x,t;\lambda),\mathbf{e}_{j})\mathbf{e}_{j}.
\end{aligned}
\end{equation}
By combining Eq.\eqref{eq:ab1} and Lemma \eqref{lemu} with Eq.\eqref{eq:admu}, we obtain
\begin{equation}
\begin{aligned}
\label{eq:sym3}
&\mu_{\mathrm{-R}}(x,t;\lambda)\mathrm{ad}[\mathbf{a}(\lambda)]\\
=&\mu_{\mathrm{-R}}(x,t;\lambda)\mathrm{ad}[(\mu_{\mathrm{+R}})^{\dag}(x,t;\bar{\lambda})\mu_{\mathrm{-R}}(x,t;\lambda)]\\
=&(\mathscr{G}_{1},\ldots,\mathscr{G}_{n})[\bar{\mu}_{\mathrm{+R}}(x,t;\bar{\lambda}),\mathcal{G}[\mu_{\mathrm{-R}}(x,t;\lambda)]]\\
=&(\mathscr{G}_{1},\ldots,\mathscr{G}_{n})[\bar{\mu}_{\mathrm{+R}}(x,t;\bar{\lambda}),(-1)^{n}[[\mathrm{ad}[\mu_{-}(x,t;\lambda)]]^{\mathrm{T}}]_{\mathrm{L}}]\\
=&(\mathscr{G}_{1},\ldots,\mathscr{G}_{n})[\bar{\mu}_{\mathrm{+R}}(x,t;\bar{\lambda}),(-1)^{n}\bar{\mu}_{\mathrm{-L}}(x,t;\bar{\lambda})].
\end{aligned}
\end{equation}
\subsection{Discrete spectrum}
\begin{prop}\label{prop:dpsi}
Suppose that $\lambda_{0}$ is a zero of $\mathrm{det}[\mathbf{a}^{\dag}(\bar{\lambda})]$ with multiplicity $m+1$. There exist $m+1$ complex-valued constant vectors $\mathbf{B}_{0},\mathbf{B}_{1},\ldots,\mathbf{B}_{m}$, where for each $s\in\left\{0,\ldots,m\right\}$, $\mathbf{B}_{s}=(\mathrm{B}_{s1},\ldots, \mathrm{B}_{sn})^{\mathrm{T}}$ and $\mathbf{B}_{0}\neq\mathbf{0}$. For each $h\in\left\{0,\ldots,m\right\}$, the following expression holds:
\begin{equation}
\label{eq:psid}
\frac{\psi_{\mathrm{-L}}^{(h)}(x,t;\lambda_{0})}{h!}=\sum_{\substack{j+k=h\\j,k\geqslant0}}\frac{\psi_{\mathrm{+R}}^{(k)}(x,t;\lambda_{0})\mathbf{B}_{j}}{j!k!}.
\end{equation}
\end{prop}
\begin{proof}When $h=0$, based on Eq.\eqref{eq:ab1}, it follows that the vectors $\psi_{-1}(x,t;\lambda_{0})$, $\psi_{+2}(x,t;\lambda_{0})$, $\ldots$, $\psi_{+(n+1)}(x,t;\lambda_{0})$ are linearly dependent. Furthermore, since the rank of $\psi_{\mathrm{+R}}(x,t;\lambda)$ is $n$, there must exist a non-zero complex-valued constant vector $\mathbf{B}_{0}$ such that $\psi_{\mathrm{-L}}(x,t;\lambda_{0})=\psi_{\mathrm{+R}}(x,t;\lambda_{0})\mathbf{B}_{0}$.

For any positive integer $j\leqslant m$, supposed that this proposition holds for all $h<j$, meaning that there exist complex-valued constant vctors $\mathbf{B}_{0},\mathbf{B}_{1},\ldots,\mathbf{B}_{j-1}$ such that for each $h\in\left\{0,\ldots,j-1\right\}$, the following relation holds:
\begin{equation}
\label{eq:ls1}
\frac{\psi_{\mathrm{-L}}^{(h)}(x,t;\lambda_{0})}{h!}=\sum_{\substack{r+s=h\\r,s\geqslant0}}\frac{\psi_{\mathrm{+R}}^{(s)}(x,t;\lambda_{0})\mathbf{B}_{r}}{r!s!}.
\end{equation}
By combining Eq.\eqref{eq:ab1} with $\frac{\mathrm{d}^{j}(\mathrm{det}[\mathbf{a}^{\dag}(\bar{\lambda})])}{\mathrm{d}\lambda^{j}}\big|_{\lambda=\lambda_{0}}=0$, we can see that
\begin{equation}
\label{eq:ls2}
\sum_{\substack{s+l_{1}+\cdots+l_{n}=j\\s,l_{1},\ldots,l_{n}\geqslant0}}\frac{j!}{s!l_{1}!\ldots l_{n}!}\mathrm{det}\left(\psi_{-1}^{(s)}(\lambda_{0}),\psi_{+2}^{(l_{1})}(\lambda_{0}),\ldots,\psi_{+(n+1)}^{(l_{n})}(\lambda_{0})\right)=0.
\end{equation}
Substituting Eq.\eqref{eq:ls1} into Eq.\eqref{eq:ls2} yields
\begin{align}
0=&\mathrm{det}\left(\psi_{-1}^{(j)}(\lambda_{0}),\psi_{+2}(\lambda_{0}),\ldots,\psi_{+(n+1)}(\lambda_{0})\right)\nonumber\\
&+\sum_{\substack{s+r+l_{1}+\cdots+l_{n}=j\\s+r\neq j\\s,l_{1},\ldots,l_{n}\geqslant0}}\frac{j!}{s!r!l_{1}!\ldots l_{n}!}\mathrm{det}\left(\psi_{+\mathrm{R}}^{(s)}(\lambda_{0})\mathbf{B}_{r},\psi_{+2}^{(l_{1})}(\lambda_{0}),\ldots,\psi_{+(n+1)}^{(l_{n})}(\lambda_{0})\right)\nonumber\\
=&\mathrm{det}\left(\psi_{\mathrm{-L}}^{(j)}(\lambda_{0}),\psi_{\mathrm{+R}}(\lambda_{0})\right)\nonumber\\
&+\Bigg(\sum_{\substack{s+r+l_{1}+\cdots+l_{n}=j\\s+r\neq j\\l_{1}+r\neq j\\\vdots\\l_{n}+r\neq j\\s,l_{1},\ldots,l_{n}\geqslant0}}+\sum_{k=1}^{n}\sum_{\substack{s+r+l_{1}+\cdots+l_{n}=j\\s+r\neq j\\l_{k}+r=j\\s,l_{1},\ldots,l_{n}\geqslant0}}-\sum_{k,m=1}^{n}\sum_{\substack{s+r+l_{1}+\cdots+l_{n}=j\\s+r\neq j\\l_{k}+r=j\\l_{m}+r=j\\s,l_{1},\ldots,l_{n}\geqslant0}}+\sum_{k,m,p=1}^{n}\sum_{\substack{s+r+l_{1}+\cdots+l_{n}=j\\s+r\neq j\\l_{k}+r=j\\l_{m}+r=j\\l_{p}+r=j\\s,l_{1},\ldots,l_{n}\geqslant0}}-\cdots\nonumber\\
&+(-1)^{n+1}\sum_{\substack{s+r+l_{1}+\cdots+l_{n}=j\\s+r\neq j\\l_{1}+r=j\\\vdots\\l_{n}+r=j\\s,l_{1},\ldots,l_{n}\geqslant0}}\Bigg)\frac{j!}{s!r!l_{1}!\ldots l_{n}!}\mathrm{det}\left(\psi_{+\mathrm{R}}^{(s)}(\lambda_{0})\mathbf{B}_{r},\psi_{+2}^{(l_{1})}(\lambda_{0}),\ldots,\psi_{+(n+1)}^{(l_{n})}(\lambda_{0})\right)\nonumber\\
=&\mathrm{det}\left(\psi_{\mathrm{-L}}^{(j)}(\lambda_{0}),\psi_{\mathrm{+R}}(\lambda_{0})\right)\nonumber\\
&+\sum_{k=1}^{n}\sum_{\substack{l_{k}+r=j\\l_{k}>0,r\geqslant0}}\frac{j!}{l_{k}!r!}\mathrm{det}\left(\psi_{\mathrm{+R}}(\lambda_{0})\mathbf{B}_{r},\psi_{+2}(\lambda_{0}),\ldots,\psi_{+(k+1)}^{(l_{k})}(\lambda_{0}),\ldots,\psi_{+(n+1)}(\lambda_{0})\right)\nonumber\\
=&\mathrm{det}\left(\psi_{\mathrm{-L}}^{(j)}(\lambda_{0}),\psi_{\mathrm{+R}}(\lambda_{0})\right)+\sum_{k=1}^{n}\sum_{\substack{l_{k}+r=j\\l_{k}>0,r\geqslant0}}\frac{j!}{l_{k}!r!}\nonumber\\
&\mathrm{det}\left(\psi_{+2}(\lambda_{0})\mathrm{B}_{r1}+\cdots+\psi_{+(n+1)}(\lambda_{0})\mathrm{B}_{rn},\psi_{+2}(\lambda_{0}),\ldots,\psi_{+(k+1)}^{(l_{k})}(\lambda_{0}),\ldots,\psi_{+(n+1)}(\lambda_{0})\right)\nonumber\\
=&\mathrm{det}\left(\psi_{\mathrm{-L}}^{(j)}(\lambda_{0}),\psi_{\mathrm{+R}}(\lambda_{0})\right)\nonumber\\
&+\sum_{k=1}^{n}\sum_{\substack{l_{k}+r=j\\l_{k}>0,r\geqslant0}}\frac{j!}{l_{k}!r!}\mathrm{det}\left(\psi_{+(k+1)}(\lambda_{0}),\psi_{+2}(\lambda_{0}),\ldots,\psi_{+(k+1)}^{(l_{k})}(\lambda_{0}),\ldots,\psi_{+(n+1)}(\lambda_{0})\right)\mathrm{B}_{rk}\nonumber\\
=&\mathrm{det}\left(\psi_{\mathrm{-L}}^{(j)}(\lambda_{0}),\psi_{\mathrm{+R}}(\lambda_{0})\right)-\sum_{\substack{l+r=j\\l>0,r\geqslant0}}\frac{j!}{l!r!}\mathrm{det}\left(\psi_{\mathrm{+R}}^{(l)}(\lambda_{0})\mathbf{B}_{r},\psi_{\mathrm{+R}}(\lambda_{0})\right)\nonumber\\
=&\mathrm{det}\bigg(\psi_{\mathrm{-L}}^{(j)}(\lambda_{0})-\sum_{\substack{l+r=j\\l>0,r\geqslant0}}\frac{j!}{l!r!}\psi_{\mathrm{+R}}^{(l)}(\lambda_{0})\mathbf{B}_{r},\psi_{\mathrm{+R}}(\lambda_{0})\bigg).
\end{align}
Since the rank of $\psi_{\mathrm{+R}}(\lambda_{0})$ is $n$, there must exist a non-zero complex-valued constant vector $\mathbf{B}_{j}$ such that
\begin{equation}
\psi_{\mathrm{-L}}^{(j)}(\lambda_{0})-\sum_{\substack{l+r=j\\l>0,r\geqslant0}}\frac{j!}{l!r!}\psi_{\mathrm{+R}}^{(l)}(\lambda_{0})\mathbf{B}_{r}=\psi_{\mathrm{+R}}(\lambda_{0})\mathbf{B}_{j}.
\end{equation}
Thus,
\begin{equation}
\psi_{\mathrm{-L}}^{(j)}(\lambda_{0})=\sum_{\substack{l+r=j\\l,r\geqslant0}}\frac{j!}{l!r!}\psi_{\mathrm{+R}}^{(l)}(\lambda_{0})\mathbf{B}_{r},
\end{equation}
which shows this proposition holds for $h=j$. Consequently, the proposition is proven.
\end{proof}
\begin{corollary}\label{eq:dcormu}
Supposed that $\lambda_{0}$ is a zero of $\mathrm{det}[\mathbf{a}^{\dag}(\bar{\lambda})]$ with multiplicity $m+1$. Consequently,  for each $h\in\left\{0,\ldots,m\right\}$, the following relationship holds:
\begin{equation}
\label{eq:muLd}
\frac{\mu_{\mathrm{-L}}^{(h)}(x,t;\lambda_{0})}{h!}=\sum_{\substack{j+k+l=h\\j,k,l\geqslant0}}\frac{\Theta^{(k)}(x,t;\lambda_{0})\mu_{\mathrm{+R}}^{(l)}(x,t;\lambda_{0})\mathbf{B}_{j}}{j!k!l!},
\end{equation}
where $\Theta(x,t;\lambda)=\mathrm{e}^{2\mathrm{i}\theta(x,t;\lambda)}$ and $\mathbf{B}_{0},\mathbf{B}_{1},\ldots,\mathbf{B}_{m}$ are given in Proposition \ref{prop:dpsi}. Moreover,
\begin{equation}
\label{eq:muRd}
\frac{[\mu_{\mathrm{-R}}(x,t;\lambda)\mathrm{ad}[\mathbf{a}(\lambda)]]^{(h)}\mid_{\lambda=\bar{\lambda}_{0}}}{h!}=-\sum_{\substack{j+k+l=h\\j,k,l\geqslant0}}\frac{\overline{\Theta^{(k)}(x,t;\lambda_{0})}\mu_{\mathrm{+L}}^{(l)}(x,t;\bar{\lambda}_{0})\mathbf{B}_{j}^{\dag}}{j!k!l!}.
\end{equation}
\end{corollary}
\begin{proof}
By combining Eq.\eqref{mJost} with Proposition \ref{prop:dpsi}, we derive
\begin{equation}
\begin{aligned}
\frac{\mu_{\mathrm{-L}}^{(h)}(\lambda_{0})}{h!}=&\frac{(\Theta^{\frac{1}{2}}\psi_{\mathrm{-L}})^{(h)}(\lambda_{0})}{h!}
=\sum_{\substack{r+s=h\\r,s\geqslant0}}\frac{(\Theta^{\frac{1}{2}})^{(r)}(\lambda_{0})\psi_{\mathrm{-L}}^{(s)}(\lambda_{0})}{r!s!}\\
=&\sum_{\substack{r+s=h\\r,s\geqslant0}}\sum_{\substack{j+m=s\\j,m\geqslant0}}\frac{(\Theta^{\frac{1}{2}})^{(r)}(\lambda_{0})\psi_{\mathrm{+R}}^{(m)}(\lambda_{0})\mathbf{B}_{j}}{r!j!m!}\\
=&\sum_{\substack{r+j+m=h\\r,j,m\geqslant0}}\frac{(\Theta^{\frac{1}{2}})^{(r)}(\lambda_{0})(\Theta^{\frac{1}{2}}\mu_{\mathrm{+R}})^{(m)}(\lambda_{0})\mathbf{B}_{j}}{r!j!m!}\\
=&\sum_{\substack{r+j+s+l=h\\r,j,s,l\geqslant0}}\frac{(\Theta^{\frac{1}{2}})^{(r)}(\lambda_{0})(\Theta^{\frac{1}{2}})^{(s)}(\lambda_{0})\mu_{\mathrm{+R}}^{(l)}(\lambda_{0})\mathbf{B}_{j}}{r!j!s!l!}\\
=&\sum_{\substack{j+k+l=h\\j,k,l\geqslant0}}\sum_{\substack{r+s=k\\r,s\geqslant0}}\frac{(\Theta^{\frac{1}{2}})^{r}(\lambda_{0})(\Theta^{\frac{1}{2}})^{(s)}(\lambda_{0})}{r!s!}\frac{\mu_{\mathrm{+R}}^{(l)}(\lambda_{0})\mathbf{B}_{j}}{j!l!}\\
=&\sum_{\substack{j+k+l=h\\j,k,l\geqslant0}}\frac{\Theta^{(k)}(\lambda_{0})\mu_{\mathrm{+R}}^{(l)}(\lambda_{0})\mathbf{B}_{j}}{j!k!l!}.
\end{aligned}
\end{equation}
It follows from Eqs.\eqref{eq:Gldef}, \eqref{eq:sym2}, \eqref{eq:sym3} and \eqref{eq:muLd} that
\begin{align}
&\frac{[\mu_{\mathrm{-R}}(\lambda)\mathrm{ad}[\mathbf{a}(\lambda)]]^{(h)}|_{\lambda=\bar{\lambda}_{0}}}{h!}\nonumber\\
=&\frac{[(\mathscr{G}_{1},\ldots,\mathscr{G}_{n})[\bar{\mu}_{\mathrm{+R}}(\bar{\lambda}),(-1)^{n}\bar{\mu}_{\mathrm{-L}}(\bar{\lambda})]]^{(h)}|_{\lambda=\bar{\lambda}_{0}}}{h!}\nonumber\\
=&\sum_{s=1}^{n}\frac{[\mathscr{G}_{s}[\bar{\mu}_{\mathrm{+R}}(\bar{\lambda}),(-1)^{n}\bar{\mu}_{\mathrm{-L}}(\bar{\lambda})]]^{(h)}|_{\lambda=\bar{\lambda}_{0}}}{h!}\mathbf{e}_{s}^{\mathrm{T}}\nonumber\\
=&(-1)^{n}\sum_{s=1}^{n}\frac{\overline{[\mathscr{G}_{s}[\mu_{\mathrm{+R}}(\lambda),\mu_{\mathrm{-L}}(\lambda)]]^{(h)}|_{\lambda=\lambda_{0}}}}{h!}\mathbf{e}_{s}^{\mathrm{T}}\nonumber\\
=&(-1)^{n}\sum_{s=1}^{n}\sum_{\substack{r_{1}+\cdots+r_{n-1}+k=h\\r_{1},\ldots,r_{n-1},k\geqslant0}}\frac{\overline{\mathscr{G}_{s}[\mu_{+2}^{(r_{1})}(\lambda_{0}),\ldots,\mu_{+(s+1)}(\lambda_{0}),\ldots,\mu_{+(n+1)}^{(r_{n-1})}(\lambda_{0}),\mu_{\mathrm{-L}}^{(k)}(\lambda_{0})]}}{r_{1}!\cdots r_{n-1}!k!}\mathbf{e}_{s}^{\mathrm{T}}\nonumber\\
=&(-1)^{n}\sum_{s=1}^{n}\sum_{\substack{r_{1}+\cdots+r_{n-1}+k+l+j=h\\r_{1},\ldots,r_{n-1},k,l,j\geqslant0}}\frac{\overline{\mathscr{G}_{s}[\mu_{+2}^{(r_{1})}(\lambda_{0}),\ldots,\mu_{+(s+1)}(\lambda_{0}),\ldots,\mu_{+(n+1)}^{(r_{n-1})}(\lambda_{0}),\Theta^{(k)}(\lambda_{0})\mu_{\mathrm{+R}}^{(l)}(\lambda_{0})\mathbf{B}_{j}]}}{r_{1}!\cdots r_{n-1}!k!l!j!}\mathbf{e}_{s}^{\mathrm{T}}\nonumber\\
=&(-1)^{n}\sum_{s=1}^{n}\sum_{\substack{r+k+j=h\\r,k,j\geqslant0}}\sum_{\substack{r_{1}+\cdots+r_{n-1}+l=r\\r_{1},\ldots,r_{n-1},l\geqslant0}}\overline{\Theta^{(k)}(\lambda_{0})}\frac{\overline{\mathscr{G}_{s}[\mu_{+2}^{(r_{1})}(\lambda_{0}),\ldots,\mu_{+(s+1)}(\lambda_{0}),\ldots,\mu_{+(n+1)}^{(r_{n-1})}(\lambda_{0}),\mu_{+(s+1)}^{(l)}(\lambda_{0})]}\bar{\mathrm{B}}_{js}}{r_{1}!\cdots r_{n-1}!k!l!j!}\mathbf{e}_{s}^{\mathrm{T}}\nonumber\\
=&(-1)^{n}\sum_{s=1}^{n}\sum_{\substack{r+k+j=h\\r,k,j\geqslant0}}\overline{\Theta^{(k)}(\lambda_{0})}\frac{\overline{[\mathscr{G}_{s}[\mu_{\mathrm{+R}}(\lambda),\mu_{+(s+1)}(\lambda)]]^{(r)}|_{\lambda=\lambda_{0}}}\bar{\mathrm{B}}_{js}}{r!j!k!}\mathbf{e}_{s}^{\mathrm{T}}\nonumber\\
=&(-1)^{n}\sum_{s=1}^{n}\sum_{\substack{r+k+j=h\\r,k,j\geqslant0}}\overline{\Theta^{(k)}(\lambda_{0})}\frac{(-1)^{n-s}\overline{[\mathscr{G}_{s}[\mu_{\mathrm{+2}}(\lambda),\ldots,\mu_{+(s+1)}(\lambda),\mu_{+(s+1)}(\lambda),\ldots,\mu_{+(n+1)}(\lambda)]]^{(r)}|_{\lambda=\lambda_{0}}}\bar{\mathrm{B}}_{js}}{r!j!k!}\mathbf{e}_{s}^{\mathrm{T}}\nonumber\\
=&(-1)^{n}\sum_{s=1}^{n}\sum_{\substack{r+k+j=h\\r,k,j\geqslant0}}\overline{\Theta^{(k)}(\lambda_{0})}\frac{(-1)^{n-s}(-1)^{n+s+1}\sum\limits_{l=1}^{n+1}\overline{[\mathrm{det}[\mu_{\mathrm{+R}}(\lambda),\mathbf{e}_{l}]]^{(r)}|_{\lambda=\lambda_{0}}}\mathbf{e}_{l}}{r!j!k!}\bar{\mathrm{B}}_{js}\mathbf{e}_{s}^{\mathrm{T}}\nonumber\\
=&(-1)^{n+1}\sum_{\substack{r+j+k=h\\r,j,k,\geqslant0}}\overline{\Theta^{(k)}(\lambda_{0})}\sum_{l=1}^{n+1}\frac{\overline{[\mathrm{det}[\mu_{\mathrm{+R}}(\lambda),\mathbf{e}_{l}]]^{(r)}|_{\lambda=\lambda_{0}}}}{r!j!k!}\mathbf{e}_{l}[\bar{\mathrm{B}}_{j1},\ldots,\bar{\mathrm{B}}_{jn}]\nonumber\\
=&-\sum_{\substack{r+j+k=h\\r,j,k,\geqslant0}}\overline{\Theta^{(k)}(\lambda_{0})}\frac{\overline{\bar{\mu}_{\mathrm{+L}}^{(r)}(\bar{\lambda})|_{\lambda=\lambda_{0}}}}{r!j!k!}\mathbf{B}_{j}^{\dag}\nonumber\\
=&-\sum_{\substack{j+k+l=h\\j,k,l\geqslant0}}\frac{\overline{\Theta^{(k)}(\lambda_{0})}\mu_{\mathrm{+L}}^{(l)}(\bar{\lambda}_{0})\mathbf{B}_{j}^{\dag}}{j!k!l!}.
\end{align}
\end{proof}
Suppose that $\lambda_{0}$ is a zero of $\mathrm{det}[\mathbf{a}^{\dag}(\bar{\lambda})]$ with multiplicity $m+1$, $\frac{1}{\mathrm{det}[\mathbf{a}^{\dag}(\bar{\lambda})]}$ can be expressed as a Laurent series expansion around $\lambda=\lambda_{0}$,
$$\frac{1}{\mathrm{det}[\mathbf{a}^{\dag}(\bar{\lambda})]}=\frac{a_{-m-1}}{(\lambda-\lambda_{0})^{m+1}}+\frac{a_{-m}}{(\lambda-\lambda_{0})^{m}}+\cdot\cdot\cdot+\frac{a_{-1}}{\lambda-\lambda_{0}}+O(1),\quad\lambda\rightarrow\lambda_{0},$$
where $a_{-m-1}\neq0$ and $a_{-h-1}=\frac{\tilde{a}^{(m-h)}(\lambda_{0})}{(m-h)!}$, with $\tilde{a}(\lambda)=\frac{(\lambda-\lambda_{0})^{m+1}}{\mathrm{det}[\mathbf{a}^{\dag}(\bar{\lambda})]}$ for $ h=0,\ldots,m$. By combining with Corollary \ref{eq:dcormu}, we can derive the following for each $h\in\left\{0,\ldots,m\right\}$:
\begin{align}
\label{res}
&\underset{\lambda=\lambda_{0}}{\mathrm{Res}}\frac{(\lambda-\lambda_{0})^{h}\mu_{\mathrm{-L}}(x,t;\lambda)}{\mathrm{det}[\mathbf{a}^{\dag}(\bar{\lambda})]}=\sum_{\substack{j+s+k+l=m-h\\j,s,k,l\geqslant0}}\frac{\tilde{a}^{(j)}(\lambda_{0})\Theta^{(l)}(x,t;\lambda_{0})\mu_{\mathrm{+R}}^{(s)}(x,t;\lambda_{0})\mathbf{B}_{k}}{j!s!k!l!},\\
\label{resc}
&\underset{\lambda=\bar{\lambda}_{0}}{\mathrm{Res}}(\lambda-\bar{\lambda}_{0})^{h}\mu_{\mathrm{-R}}(x,t;\lambda)\mathbf{a}^{-1}(\lambda)=-\sum_{\substack{j+s+k+l=m-h\\j,s,k,l\geqslant0}}\frac{\overline{\tilde{a}^{(j)}(\lambda_{0})\Theta^{(l)}(x,t;\lambda_{0})}\mu_{\mathrm{+L}}^{(s)}(x,t;\bar{\lambda}_{0})\mathbf{B}_{k}^{\dag}}{j!s!k!l!}.
\end{align}
We introduce a vector-valued polynomial $\mathbf{f}_{0}(\lambda)$ of a degree at most $m$ to encapsulate the residues more succinctly. The  polynomial $\mathbf{f}_{0}(\lambda)$ is defined as
\begin{equation}
\label{eq:f0}
\mathbf{f}_{0}(\lambda)=\sum_{l=0}^{m}\sum_{\substack{j+k=l\\j,k\geqslant0}}\frac{\tilde{a}^{(j)}(\lambda_{0})\mathbf{B}_{k}}{j!k!}(\lambda-\lambda_{0})^{l},
\end{equation}
with  the condition that $\mathbf{f}_{0}(\lambda_{0})\neq\mathbf{0}$. Consequently, we have
\begin{equation}
\begin{aligned}
\label{eq:resf0}
\underset{\lambda=\lambda_{0}}{\mathrm{Res}}\frac{(\lambda-\lambda_{0})^{h}\mu_{\mathrm{-L}}(x,t;\lambda)}{\mathrm{det}[\mathbf{a}^{\dag}(\bar{\lambda})]}
=&\sum_{\substack{r+l+s=m-h\\r,l,s\geqslant0}}\frac{\Theta^{(l)}(x,t;\lambda_{0})\mu_{\mathrm{+R}}^{(s)}(x,t;\lambda_{0})\mathbf{f}_{0}^{(r)}(\lambda_{0})}{r!l!s!}\\
=&\frac{\left[\mathrm{e}^{2\mathrm{i}\theta(x,t;\lambda)}\mu_{\mathrm{+R}}(x,t;\lambda)\mathbf{f}_{0}(\lambda)\right]^{(m-h)}\Big|_{\lambda=\lambda_{0}}}{(m-h)!},
\end{aligned}
\end{equation}
\begin{equation}
\begin{aligned}
\label{eq:rescf0}
\underset{\lambda=\bar{\lambda}_{0}}{\mathrm{Res}}(\lambda-\bar{\lambda}_{0})^{h}\mu_{\mathrm{-R}}(x,t;\lambda)\mathbf{a}^{-1}(\lambda)
=&-\sum_{\substack{r+l+s=m-h\\r,l,s\geqslant0}}\frac{\overline{\Theta^{(l)}(x,t;\lambda_{0})}\mu_{\mathrm{+L}}^{(s)}(x,t;\bar{\lambda}_{0})(\mathbf{f}_{0}^{(r)}(\lambda_{0}))^{\dag}}{r!l!s!}\\
=&-\frac{\left[\mathrm{e}^{-2\mathrm{i}\theta(x,t;\lambda)}\mu_{\mathrm{+L}}(x,t;\lambda)\mathbf{f}_{0}^{\dag}(\bar{\lambda})\right]^{(m-h)}\Big|_{\lambda=\bar{\lambda}_{0}}}{(m-h)!}.
\end{aligned}
\end{equation}
We designate $\mathbf{f}_{0}(\lambda_{0}),\ldots,\mathbf{f}_{0}^{(m)}(\lambda_{0})$ as the residue constants  associated with the discrete spectrum $\lambda_{0}$.
\begin{assume}\label{sing}
Supposed that $\mathrm{det}[\mathbf{a}^{\dag}(\bar{\lambda})]$ has $N$ pairs of distinct zeros$\left\{\lambda_{k},-\bar{\lambda}_{k}\right\}_{k=1}^{N}\in\mathbb{C^{+}}$ with multiplicity $\left\{m_{k}+1\right\}_{k=1}^{N}$, respectively. Specifically, if for $k=1,\ldots,N_{1}$, $\mathrm{Re}\lambda_{k}=0$, then $\lambda_{k}=-\bar{\lambda}_{k}$; for $k=N_{1}+1,\ldots,N$, $\mathrm{Re}\lambda_{k}\neq0$, and none of these zeros lie on the real axis.
\end{assume}
\begin{prop}\label{prop:res}
If $\lambda_{1},\ldots,\lambda_{N}$ are as described in the Assumption \ref{sing}, then there exists uniquely a vector-valued polynomial $\mathbf{f}(\lambda)$ with a degree less than $\tau=\sum\limits_{k=1}^{N}(m_{k}+1)$ such that $\mathbf{f}(\lambda_{k})\neq\mathbf{0}$, and for each $k=1,\ldots,N$, $n_{k}=0,\ldots,m_{k}$,
\begin{alignat}{2}
\label{eq:resfa}
&\underset{\lambda=\lambda_{k}}{\mathrm{Res}}\frac{(\lambda-\lambda_{k})^{n_{k}}\mu_{\mathrm{-L}}(x,t;\lambda)}{\mathrm{det}[\mathbf{a}^{\dag}(\bar{\lambda})]}=\frac{\left[\mathrm{e}^{2\mathrm{i}\theta(x,t;\lambda)}\mu_{\mathrm{+R}}(x,t;\lambda)\mathbf{f}(\lambda)\right]^{(m_{k}-n_{k})}\Big|_{\lambda=\lambda_{k}}}{(m_{k}-n_{k})!},\\
\label{eq:resfb}
&\underset{\lambda=-\bar{\lambda}_{k}}{\mathrm{Res}}\frac{(\lambda+\bar\lambda_{k})^{n_{k}}\mu_{\mathrm{-L}}(x,t;\lambda)}{\mathrm{det}[\mathbf{a}^{\dag}(\bar{\lambda})]}=\frac{\left[\mathrm{e}^{2\mathrm{i}\theta(x,t;\lambda)}\mu_{\mathrm{+R}}(x,t;\lambda)\bar{\mathbf{f}}(-\bar{\lambda})\right]^{(m_{k}-n_{k})}\Big|_{\lambda=-\bar{\lambda}_{k}}}{(-1)^{m_{k}+1}(m_{k}-n_{k})!},\\
\label{eq:resfc}
&\underset{\lambda=\bar{\lambda}_{k}}{\mathrm{Res}}(\lambda-\bar{\lambda}_{k})^{n_{k}}\mu_{\mathrm{-R}}(x,t;\lambda)\mathbf{a}^{-1}(\lambda)=-\frac{\left[\mathrm{e}^{-2\mathrm{i}\theta(x,t;\lambda)}\mu_{\mathrm{+L}}(x,t;\lambda)\mathbf{f}^{\dag}(\bar{\lambda})\right]^{(m_{k}-n_{k})}\Big|_{\lambda=\bar{\lambda}_{k}}}{(m_{k}-n_{k})!},\\
\label{eq:resfd}
&\underset{\lambda=-\lambda_{k}}{\mathrm{Res}}(\lambda+\lambda_{k})^{n_{k}}\mu_{\mathrm{-R}}(x,t;\lambda)\mathbf{a}^{-1}(\lambda)=\frac{\left[\mathrm{e}^{-2\mathrm{i}\theta(x,t;\lambda)}\mu_{\mathrm{+L}}(x,t;\lambda)\mathbf{f}^{\mathrm{T}}(-\lambda)\right]^{(m_{k}-n_{k})}\Big|_{\lambda=-\lambda_{k}}}{(-1)^{m_{k}}(m_{k}-n_{k})!}.
\end{alignat}
In addition, $\mathbf{f}(\lambda)$ satisfies the symmetry condition:
\begin{equation}
\label{eq:fsym}
\mathbf{f}^{(m_{k}-n_{k})}(\lambda_{k})=(-1)^{n_{k}+1}\overline{\mathbf{f}^{(m_{k}-n_{k})}(\lambda_{k})},\quad k=1,\ldots,N_{1}.
\end{equation}
\end{prop}
\begin{proof}Analogous to Eqs.\eqref{eq:resf0} and \eqref{eq:rescf0}, for each $k\in\left\{1,\ldots,N\right\},$ there exists a vector-valued polynomial $\mathbf{f}_{k}(\lambda)$ of degree no greater than $m_{k}$ with $\mathbf{f}_{k}(\lambda_{k})\neq\mathbf{0}$, satisfying
\begin{align}
&\underset{\lambda=\lambda_{k}}{\mathrm{Res}}\frac{(\lambda-\lambda_{k})^{n_{k}}\mu_{\mathrm{-L}}(x,t;\lambda)}{\mathrm{det}[\mathbf{a}^{\dag}(\bar{\lambda})]}=\frac{\left[\mathrm{e}^{2\mathrm{i}\theta(x,t;\lambda)}\mu_{\mathrm{+R}}(x,t;\lambda)\mathbf{f}_{k}(\lambda)\right]^{(m_{k}-n_{k})}\Big|_{\lambda=\lambda_{k}}}{(m_{k}-n_{k})!},\\
&\underset{\lambda=\bar{\lambda}_{k}}{\mathrm{Res}}(\lambda-\bar{\lambda}_{k})^{n_{k}}\mu_{\mathrm{-R}}(x,t;\lambda)\mathbf{a}^{-1}(\lambda)=-\frac{\left[\mathrm{e}^{-2\mathrm{i}\theta(x,t;\lambda)}\mu_{\mathrm{+L}}(x,t;\lambda)\mathbf{f}_{k}^{\dag}(\bar{\lambda})\right]^{(m_{k}-n_{k})}\Big|_{\lambda=\bar{\lambda}_{k}}}{(m_{k}-n_{k})!},
\end{align}
where $n_{k}=0,\ldots,m_{k}$. According to the Hermite interpolation formula, there exists a unique vector-valued polynomial $\mathbf{f}(\lambda)$ with a degree less than $\tau$ such that
\begin{eqnarray}
\label{eq:feqn}
\begin{cases}
\mathbf{f}^{(n_{1})}(\lambda_{1})=\mathbf{f}_{1}^{(n_{1})}(\lambda_{1}),&\text{$n_{1}=0,\ldots,m_{1}$},\\
{\qquad\qquad}\vdots\\
\mathbf{f}^{(n_{N})}(\lambda_{N})=\mathbf{f}_{N}^{(n_{N})}(\lambda_{N}),&\text{$n_{N}=0,\ldots,m_{N}$}.
\end{cases}
\end{eqnarray}
Thus, Eqs.\eqref{eq:resfa} and \eqref{eq:resfc} are established. From Eqs.\eqref{eq:mJostsym}, \eqref{eq:absym} and \eqref{eq:resfa}, it follows that
\begin{align}
&\underset{\lambda=-\bar{\lambda}_{k}}{\mathrm{Res}}\frac{(\lambda+\bar{\lambda}_{k})^{n_{k}}\mu_{\mathrm{-L}}(x,t;\lambda)}{\mathrm{det}[\mathbf{a}^{\dag}(\bar{\lambda})]}\nonumber\\
=&-\overline{\underset{\lambda=\lambda_{k}}{\mathrm{Res}}\left[\frac{(-\lambda+\lambda_{k})^{n_{k}}\bar{\mu}_{\mathrm{-L}}(x,t;-\bar{\lambda})}{\mathrm{det}[\mathbf{a}^{\mathrm{T}}(-\lambda)]}\right]}\nonumber\\
=&(-1)^{n_{k}+1}\overline{\underset{\lambda=\lambda_{k}}{\mathrm{Res}}\left[\frac{(\lambda-\lambda_{k})^{n_{k}}\mu_{\mathrm{-L}}(x,t;\lambda)}{\mathrm{det}[\mathbf{a}^{\dag}(\bar{\lambda})]}\right]}\\
=&(-1)^{n_{k}+1}\overline{\frac{\left[\mathrm{e}^{2\mathrm{i}\theta(x,t;\lambda)}\mu_{\mathrm{+R}}(x,t;\lambda)\mathbf{f}(\lambda)\right]^{(m_{k}-n_{k})}\Big|_{\lambda=\lambda_{k}}}{(m_{k}-n_{k})!}}\nonumber\\
=&(-1)^{n_{k}+1}(-1)^{m_{k}-n_{k}}\frac{\left[\overline{\mathrm{e}^{2\mathrm{i}\theta(x,t;-\bar{\lambda})}}\bar{\mu}_{\mathrm{+R}}(x,t;-\bar{\lambda})\bar{\mathbf{f}}(-\bar{\lambda})\right]^{(m_{k}-n_{k})}\Big|_{\lambda=-\bar{\lambda}_{k}}}{(m_{k}-n_{k})!}\nonumber\\
=&(-1)^{m_{k}+1}\frac{\left[\mathrm{e}^{2\mathrm{i}\theta(x,t;\lambda)}\mu_{\mathrm{+R}}(x,t;\lambda)\bar{\mathbf{f}}(-\bar{\lambda})\right]^{(m_{k}-n_{k})}\Big|_{\lambda=-\bar{\lambda}_{k}}}{(m_{k}-n_{k})!}.\nonumber
\end{align}
We have proven that Eq.\eqref{eq:resfb} holds. Similarly,  Eq.\eqref{eq:resfd} can be proven. For $k=1,\dots,N_{1}$, $\lambda_{k}=-\bar{\lambda}_{k}$, combining Eq.\eqref{eq:resfa} with Eq.\eqref{eq:resfb} yields Eq.\eqref{eq:fsym}.
\end{proof}
\begin{remark}
Similar to Eq.\eqref{eq:f0}, we know that the vector-valued polynomial $\mathbf{f}(\lambda)$ described in Proposition \ref{prop:res} is determined by $\det[\mathbf{a}(\lambda)]$, the set $\left\{\lambda_{k},m_{k}\right\}_{k=1}^{N}$ and a collection of nonzero complex-valued constant vectors $\left\{\mathbf{B}_{k,0},\ldots,\mathbf{B}_{k,m_{k}}\right\}_{k=1}^{N}$. If $\mathbf{f}(\lambda)$ is replaced by $\mathbf{f}(\lambda)+\prod\limits_{k=1}^{N}(\lambda-\lambda_{k})^{m_{k}+1}\mathbf{g}(\lambda)$, where $\mathbf{g}(\lambda)$ is a vector-valued function which is analytic at $\left\{\lambda_{k}\right\}_{k=1}^{N}$, then Eqs.\eqref{eq:feqn} still holds. Therefore, Proposition \ref{prop:res} can be reformulated as follows:``there exists a vector-valued function $\tilde{\mathbf{f}}(\lambda)$  that is analytic and nonzero at $\left\{\lambda_{k}\right\}_{k=1}^{N}$, and still satisfies Eqs.\eqref{eq:resfa}-\eqref{eq:resfd}".
\end{remark}
\section{Inverse problem}\label{sec:inv}
In Section 2, we outlined  the direct scattering map:
\begin{equation}
    \mathcal{D}:{\mathbf{q}(x,0)}\mapsto \left\{\gamma(\lambda),\left\{\lambda_{k},m_{k},\left\{\mathbf{f}^{(n_{k})}(\lambda_{k})\right\}_{n_{k}=0}^{m_{k}}\right\}_{k=1}^{N}\right\},
\end{equation}
linking the initial potential to its scattering data.
We now turn to the inverse scattering map:
\begin{equation}
    \mathcal{I}:\left\{\gamma(\lambda),\left\{\lambda_{k},m_{k},\left\{\mathbf{f}^{(n_{k})}(\lambda_{k})\right\}_{n_{k}=0}^{m_{k}}\right\}_{k=1}^{N}\right\}\mapsto\mathbf{q}(x,t),
\end{equation}
which reconstructs the potential from the scattering data via an $(n+1)\times (n+1)$ matrix RH problem. This process is pivotal for understanding the time evolution of solutions to the vmKdV equation.
\subsection{Riemann--Hilbert problem}
Define a piecewise meromorphic function $\mathbf{M}(x,t;\lambda)$,
\begin{equation}
\label{Mdef}
\begin{aligned}
&\mathbf{M}(x,t;\lambda)=\left(\frac{\mu_{\mathrm{-L}}(x,t;\lambda)}{\mathrm{det}[\mathbf{a}^{\dag}(\bar{\lambda})]},\mu_{\mathrm{+R}}(x,t;\lambda)\right),&\quad\lambda\in\mathbb{C}^{+},\\
&\mathbf{M}(x,t;\lambda)=\left(\mu_{\mathrm{+L}}(x,t;\lambda),\mu_{\mathrm{-R}}(x,t;\lambda)\mathbf{a}^{-1}(\lambda)\right),&\quad\lambda\in\mathbb{C}^{-}.
\end{aligned}
\end{equation}
We can utilize it to formulate an $(n+1)\times (n+1)$ matrix RH problem.
\begin{problem}\label{dfrhp1}
Seek an $(n+1)\times (n+1)$ matrix-valued function \(\mathbf{M}(x, t; \lambda)\) that satisfies the following conditions:
\begin{itemize}
\item $\mathbf{Analyticity:}$ $\mathbf{M}(x,t;\lambda)$ is analytic in $\lambda$ for $\lambda\in\mathbb{C}\backslash\left(\mathbb{R}\cup\ \left\{\pm\lambda_{k},\pm\bar{\lambda}_{k}\right\}^{N}_{k=1}\right)$;
\item $\mathbf{Normalization:}$ $\mathbf{M}(x,t;\lambda)$ has the following asymptotic behavior:
\begin{equation}
    \mathbf{M}(x,t;\lambda)\rightarrow\mathbf{I},\quad\lambda\in\mathbb{C}\setminus\mathbb{R}\rightarrow\infty;
\end{equation}
    \item $\mathbf{Jump:}$ The matrix $\mathbf{M}(x,t;\lambda)$ exhibits a jump across the oriented contour $\mathbb{R}$ expressed as
    \begin{equation}
    \label{3.1}
\mathbf{M}_{+}(x,t;\lambda)=\mathbf{M}_{-}(x,t;\lambda)\mathbf{J}(x,t;\lambda),\quad\lambda\in\mathbb{R},
\end{equation}
where $\mathbf{M}_{\pm}(x,t;\lambda)=\underset{\epsilon\rightarrow0}{\mathrm{lim}}\mathbf{M}(x,t;\lambda\pm \mathrm{i}\epsilon)$ and the jump matrix $\mathbf{J}(x,t;\lambda)$ is defined as
\begin{equation}
\label{Jdef}
\mathbf{J}(x,t;\lambda)=\begin{pmatrix}
1+\mathbf{\gamma}(\lambda)\mathbf{\gamma}^{\dag}(\bar{\lambda})&-\mathrm{e}^{-2\mathrm{i}\theta(x,t;\lambda)}\mathbf{\gamma}(\lambda)\\
-\mathrm{e}^{2\mathrm{i}\theta(x,t;\lambda)}\mathbf{\gamma}^{\dag}(\bar{\lambda})&\mathbf{I}\\
\end{pmatrix};
\end{equation}
\item $\mathbf{Residues:}$ For $k=1,\ldots,N$, $\mathbf{M}(x,t;\lambda)$ exhibits multiple poles of order $m_{k}+1$ at the points  $\lambda=\pm\lambda_{k}$ and $\lambda=\pm\bar{\lambda}_{k}$. Furthermore, the residues of $\mathbf{M}(x,t;\lambda)$ at these poles satisfy the following conditions for each $n_{k}=0,\ldots,m_{k}$:
    \begin{subequations}
    \label{Res}
\begin{align}
&\underset{\lambda=\lambda_{k}}{\mathrm{Res}}(\lambda-\lambda_{k})^{n_{k}}\mathbf{M}(x,t;\lambda)=\left(\frac{\left[\mathrm{e}^{2\mathrm{i}\theta(x,t;\lambda)}\mathbf{M}_{\mathrm{R}}(x,t;\lambda)\mathbf{f}(\lambda)\right]^{(m_{k}-n_{k})}\big|_{\lambda=\lambda_{k}}}{(m_{k}-n_{k})!},\mathbf{0}\right),\\
&\underset{\lambda=-\bar{\lambda}_{k}}{\mathrm{Res}}(\lambda+\bar{\lambda}_{k})^{n_{k}}\mathbf{M}(x,t;\lambda)=\left(\frac{\left[\mathrm{e}^{2\mathrm{i}\theta(x,t;\lambda)}\mathbf{M}_{\mathrm{R}}(x,t;\lambda)\bar{\mathbf{f}}(-\bar{\lambda})\right]^{(m_{k}-n_{k})}\big|_{\lambda=-\bar{\lambda}_{k}}}{(-1)^{m_{k}+1}(m_{k}-n_{k})!},\mathbf{0}\right),\\
&\underset{\lambda=\bar{\lambda}_{k}}{\mathrm{Res}}(\lambda-\bar{\lambda}_{k})^{n_{k}}\mathbf{M}(x,t;\lambda)=\left(\mathbf{0},-\frac{\left[\mathrm{e}^{-2\mathrm{i}\theta(x,t;\lambda)}\mathbf{M}_{\mathrm{L}}(x,t;\lambda)\mathbf{f}^{\dag}(\bar{\lambda})\right]^{(m_{k}-n_{k})}\big|_{\lambda=\bar{\lambda}_{k}}}{(m_{k}-n_{k})!}\right),\\
&\underset{\lambda=-\lambda_{k}}{\mathrm{Res}}(\lambda+\lambda_{k})^{n_{k}}\mathbf{M}(x,t;\lambda)=\left(\mathbf{0},\frac{\left[\mathrm{e}^{-2\mathrm{i}\theta(x,t;\lambda)}\mathbf{M}_{\mathrm{L}}(x,t;\lambda)\mathbf{f}^{\mathrm{T}}(-\lambda)\right]^{(m_{k}-n_{k})}\big|_{\lambda=-\lambda_{k}}}{(-1)^{m_{k}}(m_{k}-n_{k})!}\right).
\end{align}
\end{subequations}
\end{itemize}
\end{problem}
For each $k$, let $\Omega_{k}$ represent a small disk centered at $\lambda_{k}$ with a radius so small such that the disk is entirely  contained within the upper half of the complex plane, and it does not intersect with any other disks or with the set $\left\{\Omega_{-k}\right\}_{k=N_{1}+1}^{N}$, where $\Omega_{-k}=\left\{\lambda|-\lambda\in\Omega_{k}\right\}$. In addition, define $\Omega_{\pm k}^{*}=\left\{\lambda|\pm\bar{\lambda}\in\Omega_{k}\right\}$. We now introduce a new matrix-valued function $\tilde{\mathbf{M}}(x,t;\lambda)$, which is defined in relation to \(\mathbf{M}(x, t; \lambda)\) as follows:
\begin{equation}
\label{TMdef}
\tilde{\mathbf{M}}(x,t;\lambda)=
\begin{cases}
\mathbf{M}(x,t;\lambda)\mathbf{P}_{k}(x,t;\lambda),\quad&\lambda\in\Omega_{k},\quad k=1,\ldots N,\\
\mathbf{M}(x,t;\lambda)[\mathbf{P}_{k}^{\mathrm{T}}(x,t;-\lambda)]^{-1},\quad&\lambda\in\Omega_{-k},\quad k=N_{1}+1,\ldots N,\\
\mathbf{M}(x,t;\lambda)[\mathbf{P}_{k}^{\dag}(x,t;\bar{\lambda})]^{-1},\quad&\lambda\in\Omega_{k}^{*},\quad k=1,\ldots N,\\
\mathbf{M}(x,t;\lambda)\bar{\mathbf{P}}_{k}(x,t;-\bar{\lambda}),\quad&\lambda\in\Omega_{-k}^{*},\quad k=N_{1}+1,\ldots N,\\
\mathbf{M}(x,t;\lambda),\quad &\mbox{otherwise},
\end{cases}
\end{equation}
where
\begin{equation}
\label{Pdef}
\mathbf{P}_{k}(x,t;\lambda)=\begin{pmatrix}
1&\mathbf{0}\\
-\frac{\mathrm{e}^{2\mathrm{i}\theta(x,t;\lambda)}\mathbf{f}(\lambda)}{(\lambda-\lambda_{k})^{m_{k}+1}}&\mathbf{I}
\end{pmatrix}.
\end{equation}
Recognizing the matrix  $\mathbf{M}(x,t;\lambda)\mathbf{P}_{k}(x,t;\lambda)$ as having a removable singularity at $\lambda_{k}$, we proceed with the following detailed examination:
\begin{align}
&\underset{\lambda=\lambda_k}{\mathrm{Res}} {(\lambda-\lambda_{k})^{n_{k}}\tilde{\mathbf{M}}(x,t;\lambda)}\nonumber\\
=&\underset{\lambda=\lambda_k}{\mathrm{Res}} {(\lambda-\lambda_{k})^{n_{k}}\left(\mathbf{M}_{\mathrm{L}}(x,t;\lambda)-\frac{\mathrm{e}^{2\mathrm{i}\theta(x,t;\lambda)}\mathbf{M}_{\mathrm{R}}(x,t;\lambda)\mathbf{f}(\lambda)}{(\lambda-\lambda_{k})^{m_{k}+1}},\mathbf{M}_{\mathrm{R}}(x,t;\lambda)\right)}\\
=&\left(\underset{\lambda=\lambda_k}{\mathrm{Res}} {(\lambda-\lambda_{k})^{n_{k}}\mathbf{M}_{\mathrm{L}}(x,t;\lambda)}-\underset{\lambda=\lambda_k}{\mathrm{Res}}{\frac{\mathrm{e}^{2\mathrm{i}\theta(x,t;\lambda)}\mathbf{M}_{\mathrm{R}}(x,t;\lambda)\mathbf{f}(\lambda)}{(\lambda-\lambda_{k})^{m_{k}-n_{k}+1}}},\mathbf{0}\right).\nonumber
\end{align}
This result is obtained by considering the residue condition \eqref{Res} and conducting a thorough analysis of the Taylor series expansion of $\mathrm{e}^{2\mathrm{i}\theta(x,t;\lambda)}\mathbf{M}_{\mathrm{R}}(x,t;\lambda)\mathbf{f}(\lambda)$  around  $\lambda_{k}$. Consequently, it becomes apparent that for each $k$ and any $0\leqslant n_{k}\leqslant m_{k}$, we have $\underset{\lambda=\lambda_k}{\mathrm{Res}} {(\lambda-\lambda_{k})^{n_{k}}\tilde{\mathbf{M}}(x,t;\lambda)}=\mathbf{0}$. Employing a similar approach, one can demonstrate that  $\mathbf{M}(x,t;\lambda)[\mathbf{P}_{k}^{\mathrm{T}}(x,t;-\lambda)]^{-1}$ has a removable singularity at $-\lambda_{k}$,  $\mathbf{M}(x,t;\lambda)[\mathbf{P}_{k}^{\dag}(x,t;\bar{\lambda})]^{-1}$ has a removable singularity at $\bar{\lambda}_{k}$, and $\mathbf{M}(x,t;\lambda)\bar{\mathbf{P}}_{k}(x,t;-\bar{\lambda})$ has a removable singularity at $-\bar{\lambda}_{k}$. Given the definition of $\tilde{\mathbf{M}}(x,t;\lambda)$, it follows that all  points $\left\{\pm\lambda_{k},\pm\bar{\lambda}_{k}\right\}_{k=1}^{N}$ are removable singularities of $\tilde{\mathbf{M}}(x,t;\lambda)$. Therefore, it can be concluded that $\tilde{\mathbf{M}}(x,t;\lambda)$ satisfies the conditions of an equivalent RH problem that is closely related to RH Problem \ref{dfrhp1}, but with the residue conditions replaced by jump conditions defined on small contours  encircling the points $\left\{\pm\lambda_{k},\pm\bar{\lambda}_{k}\right\}_{k=1}^{N}$.
\begin{problem}\label{dfrhp2}
Seek an $(n+1)\times (n+1)$ matrix-valued function $\tilde{\mathbf{M}}(x,t;\lambda)$ that satisfies the following conditions:
\begin{itemize}
\item $\mathbf{Analyticity:}$ $\tilde{\mathbf{M}}(x,t;\lambda)$ is analytic in $\lambda$ for $\lambda\in\mathbb{C}\setminus\Sigma$, where $\Sigma=\mathbb{R}\cup\left\{\partial\Omega_{k},\partial\Omega_{ k}^{*}\right\}^{N}_{k=1} \cup\left\{\partial\Omega_{- k},\partial\Omega_{- k}^{*}\right\}^{N}_{k=N_{1}+1}$;
\item $\mathbf{Normalization:}$ $\tilde{\mathbf{M}}(x,t;\lambda)$ has the following asymptotic behavior:
$$\tilde{\mathbf{M}}(x,t;\lambda)\rightarrow\mathbf{I},\quad\lambda\in\mathbb{C}\setminus\mathbb{R}\rightarrow\infty;$$
\item $\mathbf{Jump:}$ The matrix $\tilde{\mathbf{M}}(x,t;\lambda)$ takes continuous boundary values $\tilde{\mathbf{M}}_{\pm}(x,t;\lambda)$ on $\mathbb{R}$ from the respective regions $\mathbb{C}^{\pm}$, as well as from the left and right on the contours $\left\{\partial\Omega_{k}\right\}_{k=1}^{N},\left\{\partial\Omega_{- k}\right\}_{k=N_{1}+1}^{N}$ oriented in a clockwise direction and $\left\{\partial\Omega_{k}^{*}\right\}_{k=1}^{N},\left\{\partial\Omega_{-k}^{*}\right\}_{k=N_{1}+1}^{N}$ oriented in counterclockwise direction. These boundary values are interconnected through specific jump conditions,
\begin{equation}
\label{eq:TMjump}
\tilde{\mathbf{M}}_{+}(x,t;\lambda)=\tilde{\mathbf{M}}_{-}(x,t;\lambda)\tilde{\mathbf{J}}(x,t;\lambda),\quad\lambda\in\Sigma,
\end{equation}
where
\begin{equation}
\label{TJdef}
\tilde{\mathbf{J}}(x,t;\lambda)=
\begin{cases}
\mathbf{J}(x,t;\lambda),\quad &\lambda\in\mathbb{R},\\
\mathbf{P}_{k}^{-1}(x,t;\lambda),\quad&\lambda\in\partial\Omega_{k},\quad k=1,\ldots N,\\
\mathbf{P}_{k}^{\mathrm{T}}(x,t;-\lambda),\quad&\lambda\in\partial\Omega_{-k},\quad k=N_{1}+1,\ldots N,\\
[\mathbf{P}_{k}^{\dag}(x,t;\bar{\lambda})]^{-1},\quad&\lambda\in\partial\Omega_{k}^{*},\quad k=1,\ldots N,\\
\bar{\mathbf{P}}_{k}(x,t;-\bar{\lambda}),\quad&\lambda\in\partial\Omega_{-k}^{*},\quad k=N_{1}+1,\ldots N.\\
\end{cases}
\end{equation}
\end{itemize}
\end{problem}
The unique existence of the solution to RH problem \ref{dfrhp2} is contingent upon the validity of the following lemma, which serves as a foundational component of our analysis (refer to Theorem 9.3 in Ref.\cite{Zhou1989})
\begin{lemma}\label{VL}(Vanishing Lemma)If the asymptotic condition in the RH problem \ref{dfrhp2} for $\tilde{\mathbf{M}}(x,t;\lambda)$ is replaced by
\begin{equation}
\tilde{\mathbf{M}}(x,t;\lambda)=O(\lambda^{-1}),\quad\lambda\rightarrow\infty,
\end{equation}
then the RH problem \ref{dfrhp2} has only the trivial solution.
\end{lemma}
\begin{proof}Consider the function
\begin{equation}
\label{Hdef}
\mathbf{H}(x,t;\lambda)=\tilde{\mathbf{M}}(x,t;\lambda)\tilde{\mathbf{M}}^{\dag}(x,t;\bar{\lambda}),
\end{equation}
where $\mathbf{H}(x,t;\lambda)$ is  analytic in $\lambda$ for $\lambda\in\mathbb{C}\setminus\Sigma$, and continuous up to $\Sigma$. The jump of $\mathbf{H}(x,t;\lambda)$ across $\lambda\in\Sigma$ is derived as follows. Note that
\begin{equation}
\mathbf{H}_{+}(x,t;\lambda)=\tilde{\mathbf{M}}_{+}(x,t;\lambda)\tilde{\mathbf{M}}_{-}^{\dag}(x,t;\bar{\lambda}),
\end{equation}
for instance, if $\lambda$ approaches $\partial\Omega_{k}$ from the left(“+”side), then $\bar{\lambda}$ approaches $\partial\Omega_{k}^{*}$ from the right(“--”side). Here, the $``\pm"$ subscripts denote boundary values on $\partial\Omega_{k}$ $(for \lambda)$ and $\partial\Omega_{k}^{*}$ $(for \bar{\lambda})$, respectively. Applying the jump conditions across $\Sigma$, we obtain
\begin{equation}
\label{HJump}
\mathbf{H}_{+}(x,t;\lambda)=\tilde{\mathbf{M}}_{+}(x,t;\lambda)\tilde{\mathbf{M}}_{-}^{\dag}(x,t;\bar{\lambda})=\tilde{\mathbf{M}}_{-}(x,t;\lambda)\tilde{\mathbf{J}}(x,t;\lambda)\tilde{\mathbf{M}}_{-}^{\dag}(x,t;\bar{\lambda}),
\end{equation}
Using the property $\tilde{\mathbf{J}}(x,t;\lambda)=\tilde{\mathbf{J}}^{\dag}(x,t;\bar{\lambda})$, we find
\begin{equation}
\mathbf{H}_{+}(x,t;\lambda)=\tilde{\mathbf{M}}_{-}(x,t;\lambda)\tilde{\mathbf{J}}^{\dag}(x,t;\bar{\lambda})\tilde{\mathbf{M}}_{-}^{\dag}(x,t;\bar{\lambda})=\tilde{\mathbf{M}}_{-}(x,t;\lambda)\tilde{\mathbf{M}}_{+}^{\dag}(x,t;\bar{\lambda})=\mathbf{H}_{-}(x,t;\lambda).
\end{equation}
This implies that $\mathbf{H}(x,t;\lambda)$ is continuous across the entire complex $\lambda-$plane. By Morera's theorem, $\mathbf{H}(x,t;\lambda)$ is an entire function of $\lambda$. Given that $\underset{\lambda\rightarrow\infty}{\mathrm{lim}}\mathbf{H}(x,t;\lambda)=\mathbf{0}$, Liouville's theorem implies
\begin{equation}
\mathbf{H}(x,t;\lambda)\equiv\mathbf{0}.
\end{equation}
Since $\tilde{\mathbf{J}}(x,t;\lambda)$ is positive definite for $\lambda\in\mathbb{R}$, it follows directly from Eq.\eqref{HJump} that $\tilde{\mathbf{M}}_{-}(x,t;\lambda)=\mathbf{0}$ holds for $\lambda\in\mathbb{R}$. From the jump condition, it can be concluded that $\tilde{\mathbf{M}}_{+}(x,t;\lambda)=\mathbf{0}$. By analytic continuation, it follows that $\tilde{\mathbf{M}}(x,t;\lambda)=\mathbf{0}$ holds identically, extending to the boundaries $\left\{\partial\Omega_{k},\partial\Omega_{ k}^{*}\right\}^{N}_{k=1}$ and $\left\{\partial\Omega_{- k},\partial\Omega_{- k}^{*}\right\}^{N}_{k=N_{1}+1}$. Applying the jump condition for $\tilde{\mathbf{M}}(x,t;\lambda)$ along these arcs again confirms that $\tilde{\mathbf{M}}(x,t;\lambda)=\mathbf{0}$ holds for $\lambda$ within the interior of $\left\{\Omega_{k},\Omega_{ k}^{*}\right\}^{N}_{k=1}$ and $\left\{\Omega_{- k},\Omega_{- k}^{*}\right\}^{N}_{k=N_{1}+1}$. Consequently, $\tilde{\mathbf{M}}(x,t;\lambda)=\mathbf{0}$ holds throughout the entire complex plane.
\end{proof}
We now present a theorem that establishes the relationship between the solution of  RH problem \ref{dfrhp2} and the solution to vmKdV  equation \eqref{eq:vmKdVe}.
\begin{theorem}If $\tilde{\mathbf{M}}(x,t;\lambda)$ is the solution of RH problem \ref{dfrhp2}, then
\begin{equation}
\label{eq:qsym}
\mathbf{q}(x,t)=-2\mathrm{i}\underset{\lambda\rightarrow\infty}{\mathrm{lim}}\lambda\tilde{\mathbf{M}}_{\mathrm{DL}}(x,t;\lambda),
\end{equation}
is a solution to the vmKdV equation \eqref{eq:vmKdVe}.
\end{theorem}
\begin{proof}
This result is a consequence of the dressing method, as detailed in Ref.\cite{Fokas2008}.
\end{proof}
From the jump condition \eqref{eq:TMjump}, it immediately follows that for $\lambda\in\Sigma$,
\begin{equation}
\tilde{\mathbf{M}}_{+}(x,t;\lambda)-\tilde{\mathbf{M}}_{-}(x,t;\lambda)=\tilde{\mathbf{M}}_{-}(x,t;\lambda)[\tilde{\mathbf{J}}(x,t;\lambda)-\mathbf{I}]=\tilde{\mathbf{M}}_{+}(x,t;\lambda)[\mathbf{I}-\tilde{\mathbf{J}}^{-1}(x,t;\lambda)].
\end{equation}
By applying the Sokhotski--Plemelj formula, $\tilde{\mathbf{M}}(x,t;\lambda)$ can be represented as an integral:
\begin{equation}
\label{eq:TMint}
\begin{aligned}
\tilde{\mathbf{M}}(x,t;\lambda)=\mathbf{I}&+\frac{1}{2\pi\mathrm{i}}\int_{\mathbb{R}}\frac{\tilde{\mathbf{M}}_{-}(x,t;\mu)[\mathbf{J}(x,t;\mu)-\mathbf{I}]}{\mu-\lambda}\mathrm{d}\mu\\
&+\sum_{k=1}^{N}\frac{1}{2\pi\mathrm{i}}\int_{\partial\Omega_{k}}\frac{\tilde{\mathbf{M}}_{-}(x,t;\mu)[\mathbf{P}_{k}^{-1}(x,t;\mu)-\mathbf{I}]}{\mu-\lambda}\mathrm{d}\mu\\
&+\sum_{k=N_{1}+1}^{N}\frac{1}{2\pi\mathrm{i}}\int_{\partial\Omega_{-k}}\frac{\tilde{\mathbf{M}}_{-}(x,t;\mu)[\mathbf{P}_{k}^{\mathrm{T}}(x,t;-\mu)-\mathbf{I}]}{\mu-\lambda}\mathrm{d}\mu\\
&+\sum_{k=1}^{N}\frac{1}{2\pi\mathrm{i}}\int_{\partial\Omega_{k}^{*}}\frac{\tilde{\mathbf{M}}_{+}(x,t;\mu)[\mathbf{I}-\mathbf{P}_{k}^{\dag}(x,t;\bar{\mu})]}{\mu-\lambda}\mathrm{d}\mu\\
&+\sum_{k=N_{1}+1}^{N}\frac{1}{2\pi\mathrm{i}}\int_{\partial\Omega_{-k}^{*}}\frac{\tilde{\mathbf{M}}_{+}(x,t;\mu)[\mathbf{I}-(\bar{\mathbf{P}}_{k}(x,t;-\bar{\mu}))^{-1}]}{\mu-\lambda}\mathrm{d}\mu.
\end{aligned}
\end{equation}
\subsection{Reflectionless potential}
The potential $\mathbf{q}(x,t)$ is now explicitly reconstructed in the reflectionless case, where $\gamma(\lambda)=\mathbf{0}$. In this scenario, there is no jump across the contour $\mathbb{R}$, then Eq.\eqref{eq:TMint} becomes the following expression
\begin{equation}
\begin{aligned}
\tilde{\mathbf{M}}(x,t;\lambda)=\mathbf{I}&+\sum_{k=1}^{N}\frac{1}{2\pi\mathrm{i}}\int_{\partial\Omega_{k}}\frac{\tilde{\mathbf{M}}_{-}(x,t;\mu)[\mathbf{P}_{k}^{-1}(x,t;\mu)-\mathbf{I}]}{\mu-\lambda}\mathrm{d}\mu\\
&+\sum_{k=N_{1}+1}^{N}\frac{1}{2\pi\mathrm{i}}\int_{\partial\Omega_{-k}}\frac{\tilde{\mathbf{M}}_{-}(x,t;\mu)[\mathbf{P}_{k}^{\mathrm{T}}(x,t;-\mu)-\mathbf{I}]}{\mu-\lambda}\mathrm{d}\mu\\
&+\sum_{k=1}^{N}\frac{1}{2\pi\mathrm{i}}\int_{\partial\Omega_{k}^{*}}\frac{\tilde{\mathbf{M}}_{+}(x,t;\mu)[\mathbf{I}-\mathbf{P}_{k}^{\dag}(x,t;\bar{\mu})]}{\mu-\lambda}\mathrm{d}\mu\\
&+\sum_{k=N_{1}+1}^{N}\frac{1}{2\pi\mathrm{i}}\int_{\partial\Omega_{-k}^{*}}\frac{\tilde{\mathbf{M}}_{+}(x,t;\mu)[\mathbf{I}-(\bar{\mathbf{P}}_{k}(x,t;-\bar{\mu}))^{-1}]}{\mu-\lambda}\mathrm{d}\mu.
\end{aligned}
\end{equation}
Using Cauchy's Residue theorem and leveraging the definition of $\mathbf{P}_{k}(x,t;\lambda)$, the inverse problem is simplified to an algebraic system:
\begin{equation}
\begin{aligned}
\tilde{\mathbf{M}}_{\mathrm{L}}(x,t;\lambda)=&
\begin{pmatrix} 1\\\mathbf{0}\end{pmatrix}
-\sum_{k=1}^{N}\underset{\mu=\lambda_{k}}{\mathrm{Res}}{\frac{\mathrm{e}^{2\mathrm{i}\theta(x,t;\mu)}\tilde{\mathbf{M}}_{\mathrm{R}}(x,t;\mu)\mathbf{f}(\mu)}{(\mu-\lambda)(\mu-\lambda_{k})^{m_{k}+1}}}\\
&+\sum_{k=N_{1}+1}^{N}\underset{\mu=-\bar{\lambda}_{k}}{\mathrm{Res}}{(-1)^{m_{k}}\frac{\mathrm{e}^{2\mathrm{i}\theta(x,t;\mu)}\tilde{\mathbf{M}}_{\mathrm{R}}(x,t;\mu)\bar{\mathbf{f}}(-\bar{\mu})}{(\mu-\lambda)(\mu+\bar{\lambda}_{k})^{m_{k}+1}}}\\
=&
\begin{pmatrix} 1\\\mathbf{0}\end{pmatrix}
-\sum_{k=1}^{N}\frac{1}{m_{k}!}\partial_{\mu}^{m_{k}}\left(\frac{\mathrm{e}^{2\mathrm{i}\theta(x,t;\mu)}\tilde{\mathbf{M}}_{\mathrm{R}}(x,t;\mu)\mathbf{f}(\mu)}{\mu-\lambda}\right)\bigg|_{\mu=\lambda_{k}}\\
&+\sum_{k=N_{1}+1}^{N}\frac{(-1)^{m_{k}}}{m_{k}!}\partial_{\mu}^{m_{k}}\left(\frac{\mathrm{e}^{2\mathrm{i}\theta(x,t;\mu)}\tilde{\mathbf{M}}_{\mathrm{R}}(x,t;\mu)\bar{\mathbf{f}}(-\bar{\mu})}{\mu-\lambda}\right)\bigg|_{\mu=-\bar{\lambda}_{k}}\\
=&
\begin{pmatrix} 1\\\mathbf{0}\end{pmatrix}
+\sum_{k=1}^{N}\frac{1}{m_{k}!}\partial_{\mu}^{m_{k}}\left(\frac{\mathrm{e}^{2\mathrm{i}\theta(x,t;\mu)}\tilde{\mathbf{M}}_{\mathrm{R}}(x,t;\mu)\mathbf{f}(\mu)}{\lambda-\mu}\right)\bigg|_{\mu=\lambda_{k}}\\
&-\sum_{k=N_{1}+1}^{N}\frac{1}{m_{k}!}\partial_{\mu}^{m_{k}}\left(\frac{\mathrm{e}^{-2\mathrm{i}\theta(x,t;\mu)}\tilde{\mathbf{M}}_{\mathrm{R}}(x,t;-\mu)\bar{\mathbf{f}}(\bar{\mu})}{\mu+\lambda}\right)\bigg|_{\mu=\bar{\lambda}_{k}},
\end{aligned}
\end{equation}
\begin{equation}
\begin{aligned}
\tilde{\mathbf{M}}_{\mathrm{R}}(x,t;\lambda)=&
\begin{pmatrix} \mathbf{0}\\\mathbf{I}\end{pmatrix}
+\sum_{k=1}^{N}\underset{\mu=\bar{\lambda}_{k}}{\mathrm{Res}}{\frac{\mathrm{e}^{-2\mathrm{i}\theta(x,t;\mu)}\tilde{\mathbf{M}}_{\mathrm{L}}(x,t;\mu)\mathbf{f}^{\dag}(\bar{\mu})}{(\mu-\lambda)(\mu-\bar{\lambda}_{k})^{m_{k}+1}}}\\
&-\sum_{k=N_{1}+1}^{N}\underset{\mu=-\lambda_{k}}{\mathrm{Res}}{(-1)^{m_{k}}\frac{\mathrm{e}^{-2\mathrm{i}\theta(x,t;\mu)}\tilde{\mathbf{M}}_{\mathrm{L}}(x,t;\mu)\mathbf{f}^{\mathrm{T}}(-\mu)}{(\mu-\lambda)(\mu+\lambda_{k})^{m_{k}+1}}}\\
=&
\begin{pmatrix} \mathbf{0}\\\mathbf{I}\end{pmatrix}
+\sum_{k=1}^{N}\frac{1}{m_{k}!}\partial_{\mu}^{m_{k}}\left(\frac{\mathrm{e}^{-2\mathrm{i}\theta(x,t;\mu)}\tilde{\mathbf{M}}_{\mathrm{L}}(x,t;\mu)\mathbf{f}^{\dag}(\bar{\mu})}{\mu-\lambda}\right)\bigg|_{\mu=\bar{\lambda}_{k}}\\
&-\sum_{k=N_{1}+1}^{N}\frac{(-1)^{m_{k}}}{m_{k}!}\partial_{\mu}^{m_{k}}\left(\frac{\mathrm{e}^{-2\mathrm{i}\theta(x,t;\mu)}\tilde{\mathbf{M}}_{\mathrm{L}}(x,t;\mu)\mathbf{f}^{\mathrm{T}}(-\mu)}{\mu-\lambda}\right)\bigg|_{\mu=-\lambda_{k}}\\
=&
\begin{pmatrix} \mathbf{0}\\\mathbf{I}\end{pmatrix}
+\sum_{k=1}^{N}\frac{1}{m_{k}!}\partial_{\mu}^{m_{k}}\left(\frac{\mathrm{e}^{-2\mathrm{i}\theta(x,t;\mu)}\tilde{\mathbf{M}}_{\mathrm{L}}(x,t;\mu)\mathbf{f}^{\dag}(\bar{\mu})}{\mu-\lambda}\right)\bigg|_{\mu=\bar{\lambda}_{k}}\\
&+\sum_{k=N_{1}+1}^{N}\frac{1}{m_{k}!}\partial_{\mu}^{m_{k}}\left(\frac{\mathrm{e}^{2\mathrm{i}\theta(x,t;\mu)}\tilde{\mathbf{M}}_{\mathrm{L}}(x,t;-\mu)\mathbf{f}^{\mathrm{T}}(\mu)}{\mu+\lambda}\right)\bigg|_{\mu=\lambda_{k}}.
\end{aligned}
\end{equation}
More precisely,
\begin{subequations}
\begin{align}
\label{eq:TMDR}
\tilde{\mathbf{M}}_{\mathrm{\mathrm{DR}}}(x,t;\lambda)=&\mathbf{I}+\sum_{k=1}^{N}\frac{1}{m_{k}!}\partial_{\mu}^{m_{k}}\left(\frac{\mathrm{e}^{-2\mathrm{i}\theta(x,t;\mu)}\tilde{\mathbf{M}}_{\mathrm{DL}}(x,t;\mu)\mathbf{f}^{\dag}(\bar{\mu})}{\mu-\lambda}\right)\bigg|_{\mu=\bar{\lambda}_{k}}\nonumber\\
&+\sum_{k=N_{1}+1}^{N}\frac{1}{m_{k}!}\partial_{\mu}^{m_{k}}\left(\frac{\mathrm{e}^{2\mathrm{i}\theta(x,t;\mu)}\tilde{\mathbf{M}}_{\mathrm{\mathrm{DL}}}(x,t;-\mu)\mathbf{f}^{\mathrm{T}}(\mu)}{\lambda+\mu}\right)\bigg|_{\mu=\lambda_{k}},\\
\label{eq:TMDL}
\tilde{\mathbf{M}}_{\mathrm{\mathrm{DL}}}(x,t;\mu)=&\sum_{l=1}^{N}\frac{1}{m_{l}!}\partial_{\nu}^{m_{l}}\left(\frac{\mathrm{e}^{2\mathrm{i}\theta(x,t;\nu)}\tilde{\mathbf{M}}_{\mathrm{DR}}(x,t;\nu)\mathbf{f}(\nu)}{\mu-\nu}\right)\bigg|_{\nu=\lambda_{l}}\nonumber\\
&-\sum_{l=N_{1}+1}^{N}\frac{1}{m_{l}!}\partial_{\mu}^{m_{l}}\left(\frac{\mathrm{e}^{-2\mathrm{i}\theta(x,t;\nu)}\tilde{\mathbf{M}}_{\mathrm{DR}}(x,t;-\nu)\bar{\mathbf{f}}(\bar{\nu})}{\mu+\nu}\right)\bigg|_{\nu=\bar{\lambda}_{l}}.
\end{align}
\end{subequations}
Let
\begin{align}
&\mathbf{h}(\lambda)=\mathrm{e}^{2\mathrm{i}\theta(\lambda)}\mathbf{f}(\lambda),\label{hdef}\\
&\mathbf{F}_{1}(\lambda)=-2\mathrm{i}\tilde{\mathbf{M}}_{\mathrm{DR}}(\lambda)\mathbf{h}(\lambda),\\
&\mathbf{F}_{2}(\lambda)=2\mathrm{i}\tilde{\mathbf{M}}_{\mathrm{DR}}(-\lambda)\bar{\mathbf{h}}(\bar{\lambda}),
\end{align}
\begin{align}
\label{G1def}
\mathbf{G}_{1}(\lambda)=&\mathbf{F}_{1}(\lambda)+2\mathrm{i}\mathbf{h}(\lambda)+\sum_{k=1}^{N}\sum_{l=1}^{N}\frac{\partial_{\mu}^{m_{k}}\partial_{\nu}^{m_{l}}}{m_{k}!m_{l}!}\left(\frac{\mathbf{F}_{1}(\nu)\mathbf{h}^{\dag}(\bar{\mu})\mathbf{h}(\lambda)}{(\lambda-\mu)(\mu-\nu)}\right)\bigg|_{\substack{\mu=\bar{\lambda}_{k}\\ \nu=\lambda_{l}}}\nonumber\\
&+\sum_{k=1}^{N}\sum_{l=N_{1}+1}^{N}\frac{\partial_{\mu}^{m_{k}}\partial_{\nu}^{m_{l}}}{m_{k}!m_{l}!}\left(\frac{\mathbf{F}_{2}(\nu)\mathbf{h}^{\dag}(\bar{\mu})\mathbf{h}(\lambda)}{(\lambda-\mu)(\mu+\nu)}\right)\bigg|_{\substack{\mu=\bar{\lambda}_{k}\\ \nu=\bar{\lambda}_{l}}}\nonumber\\
&+\sum_{k=N_{1}+1}^{N}\sum_{l=1}^{N}\frac{\partial_{\mu}^{m_{k}}\partial_{\nu}^{m_{l}}}{m_{k}!m_{l}!}\left(\frac{\mathbf{F}_{1}(\nu)\mathbf{h}^{\mathrm{T}}(\mu)\mathbf{h}(\lambda)}{(\lambda+\mu)(\mu+\nu)}\right)\bigg|_{\substack{\mu=\lambda_{k}\\ \nu=\lambda_{l}}}\\
&+\sum_{k=N_{1}+1}^{N}\sum_{l=N_{1}+1}^{N}\frac{\partial_{\mu}^{m_{k}}\partial_{\nu}^{m_{l}}}{m_{k}!m_{l}!}\left(\frac{\mathbf{F}_{2}(\nu)\mathbf{h}^{\mathrm{T}}(\mu)\mathbf{h}(\lambda)}{(\lambda+\mu)(\mu-\nu)}\right)\bigg|_{\substack{\mu=\lambda_{k}\\ \nu=\bar{\lambda}_{l}}},\nonumber
\end{align}
\begin{align}
\label{G2def}
\mathbf{G}_{2}(\lambda)=&\mathbf{F}_{2}(\lambda)-2\mathrm{i}\bar{\mathbf{h}}(\bar{\lambda})+\sum_{k=1}^{N}\sum_{l=1}^{N}\frac{\partial_{\mu}^{m_{k}}\partial_{\nu}^{m_{l}}}{m_{k}!m_{l}!}\left(\frac{\mathbf{F}_{1}(\nu)\mathbf{h}^{\dag}(\bar{\mu})\bar{\mathbf{h}}(\bar{\lambda})}{(\lambda+\mu)(\mu-\nu)}\right)\bigg|_{\substack{\mu=\bar{\lambda}_{k}\\ \nu=\lambda_{l}}}\nonumber\\
&+\sum_{k=1}^{N}\sum_{l=N_{1}+1}^{N}\frac{\partial_{\mu}^{m_{k}}\partial_{\nu}^{m_{l}}}{m_{k}!m_{l}!}\left(\frac{\mathbf{F}_{2}(\nu)\mathbf{h}^{\dag}(\bar{\mu})\bar{\mathbf{h}}(\bar{\lambda})}{(\lambda+\mu)(\mu+\nu)}\right)\bigg|_{\substack{\mu=\bar{\lambda}_{k}\\ \nu=\bar{\lambda}_{l}}}\nonumber\\
&+\sum_{k=N_{1}+1}^{N}\sum_{l=1}^{N}\frac{\partial_{\mu}^{m_{k}}\partial_{\nu}^{m_{l}}}{m_{k}!m_{l}!}\left(\frac{\mathbf{F}_{1}(\nu)\mathbf{h}^{\mathrm{T}}(\mu)\bar{\mathbf{h}}(\bar{\lambda})}{(\lambda-\mu)(\mu+\nu)}\right)\bigg|_{\substack{\mu=\lambda_{k}\\ \nu=\lambda_{l}}}\\
&+\sum_{k=N_{1}+1}^{N}\sum_{l=N_{1}+1}^{N}\frac{\partial_{\mu}^{m_{k}}\partial_{\nu}^{m_{l}}}{m_{k}!m_{l}!}\left(\frac{\mathbf{F}_{2}(\nu)\mathbf{h}^{\mathrm{T}}(\mu)\bar{\mathbf{h}}(\bar{\lambda})}{(\lambda-\mu)(\mu-\nu)}\right)\bigg|_{\substack{\mu=\lambda_{k}\\ \nu=\bar{\lambda}_{l}}}.\nonumber
\end{align}
Substituting Eq.\eqref{eq:TMDL} into Eq.\eqref{eq:TMDR}, we can conclude that
\begin{align}
&\mathbf{G}_{1}(x,t;\lambda)=\mathbf{0},\quad(x,t;\lambda)\in\mathbb{R}\times\mathbb{R^{+}}\times\mathbb{C},  \lambda\neq\bar{\lambda}_{1},\ldots,\bar{\lambda}_{N},-\lambda_{N_{1}+1},\ldots,-\lambda_{N},\\
&\mathbf{G}_{2}(x,t;\lambda)=\mathbf{0},\quad(x,t;\lambda)\in\mathbb{R}\times\mathbb{R^{+}}\times\mathbb{C},\lambda\neq-\bar{\lambda}_{1},\ldots,-\bar{\lambda}_{N},\lambda_{N_{1}+1},\ldots,\lambda_{N}.
\end{align}
\begin{theorem}In the reflectionless case, the solution to the vmKdV equation \eqref{eq:vmKdVe} can be articulated as:
\begin{equation}
\mathbf{q}(x,t)=\sum_{k=1}^{N}\frac{\mathbf{F}_{1}^{(m_{k})}(x,t;\lambda_{k})}{m_{k}!}+\sum_{k=N_{1}+1}^{N}\frac{\mathbf{F}_{2}^{(m_{k})}(x,t;\bar{\lambda}_{k})}{m_{k}!},
\end{equation}
where the set $\bigg\{\Big\{\mathbf{F}_{1}^{(j_{k})}(x,t;\lambda_{k})\Big\}_{\substack{k=1,\ldots,N\\j_{k}=0,\ldots,m_{k}}},\Big\{\mathbf{F}_{2}^{(j_{k})}(x,t;\bar{\lambda}_{k})\Big\}_{\substack{k=N_{1}+1,\ldots,N\\j_{k}=0,\ldots,m_{k}}}\bigg\}$ denotes the solution to the ensuing algebraic system
\begin{eqnarray}
\label{algesys}
\begin{cases}
\frac{\mathbf{G}_{1}^{(j_{1})}(x,t;\lambda_{1})}{j_{1}!}=\mathbf{0},&\text{$j_{1}=0,1,\ldots,m_{1}$},\\
{\qquad}\vdots\\
\frac{\mathbf{G}_{1}^{(j_{N})}(x,t;\lambda_{N})}{j_{N}!}=\mathbf{0},&\text{$j_{N}=0,1,\ldots,m_{N}$},\\
\frac{\mathbf{G}_{2}^{(j_{N_{1}+1})}(x,t;\bar{\lambda}_{N_{1}+1})}{j_{N_{1}+1}!}=\mathbf{0},&\text{$j_{N_{1}+1}=0,1,\ldots,m_{N_{1}+1}$},\\
{\qquad}\vdots\\
\frac{\mathbf{G}_{2}^{(j_{N})}(x,t;\bar{\lambda}_{N})}{j_{N}!}=\mathbf{0},&\text{$j_{N}=0,1,\ldots,m_{N}$}.\\
\end{cases}
\end{eqnarray}
\end{theorem}
\subsection{Some explicit solutions}
In the subsequent analysis, we will delve into the numerical characteristics of the multi-pole soliton solutions for the vmKdV equation \eqref{eq:vmKdVe} in a three-component system. This exploration will involve selecting various parameter values to understand how they influence the behavior and properties of the solutions. By examining different scenarios with distinct parameter sets, we aim to uncover the richness and complexity of the solutions to this nonlinear partial differential equation.  Our goal is to gain a deeper understanding of the vmKdV equation's solutions and their implications in the context of multi-component systems.

\begin{figure}[!htbp]
\centering
 \includegraphics[scale=.5]{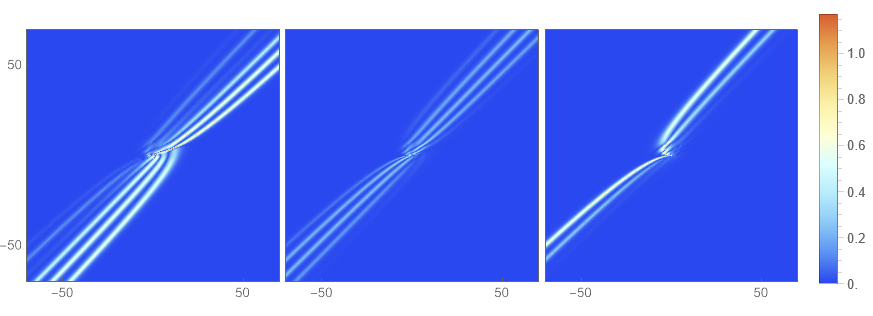}
\caption{ \label{QD3115}Density structures of three components $|q_1|$, $|q_2|$ and $ |q_{3}|$ (from left to right) as $N=1$, $\lambda_1=\frac{\mathrm{i}}{2}$, $m_1=5$, $\mathbf{f}(\lambda)=\left(1, (\lambda-\lambda_1)^2,
  (\lambda-\lambda_1)^4\right)^\mathrm{T}$.}
\end{figure}

\begin{figure}[!htbp]
\centering
 \includegraphics[scale=.5]{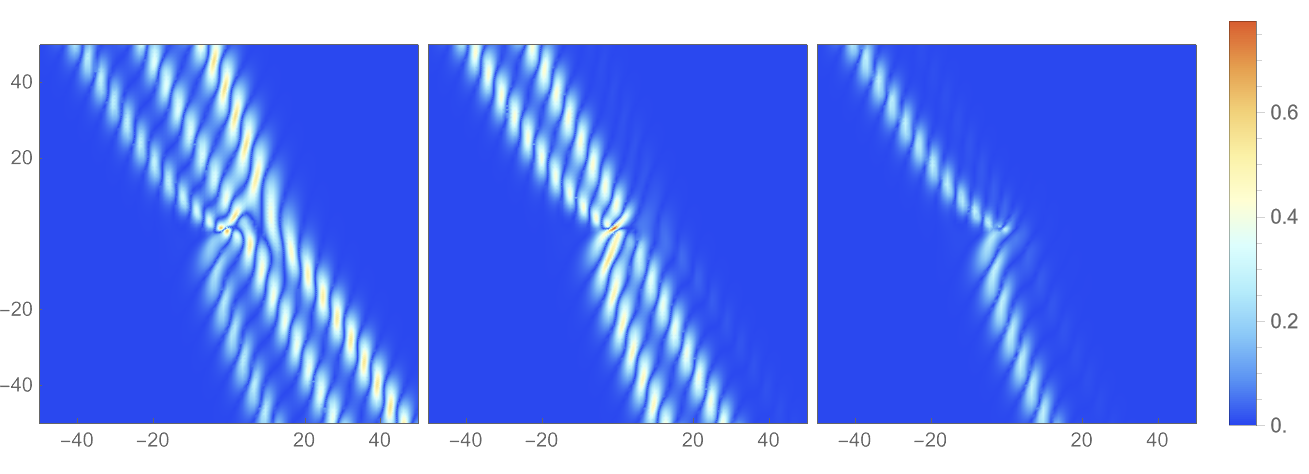}
\caption{ \label{QD3102}Density structures of three components $|q_1|$, $|q_2|$ and $ |q_{3}|$ (from left to right) as $N=1$, $\lambda_1=\frac{1+\mathrm{i}}{4}$, $m_1=2$, $\mathbf{f}(\lambda)=\left(1,\lambda-\lambda_1,(\lambda-\lambda_1)^2\right)^\mathrm{T}$.}
\end{figure}

\begin{figure}[!htbp]
\centering
 \includegraphics[scale=.5]{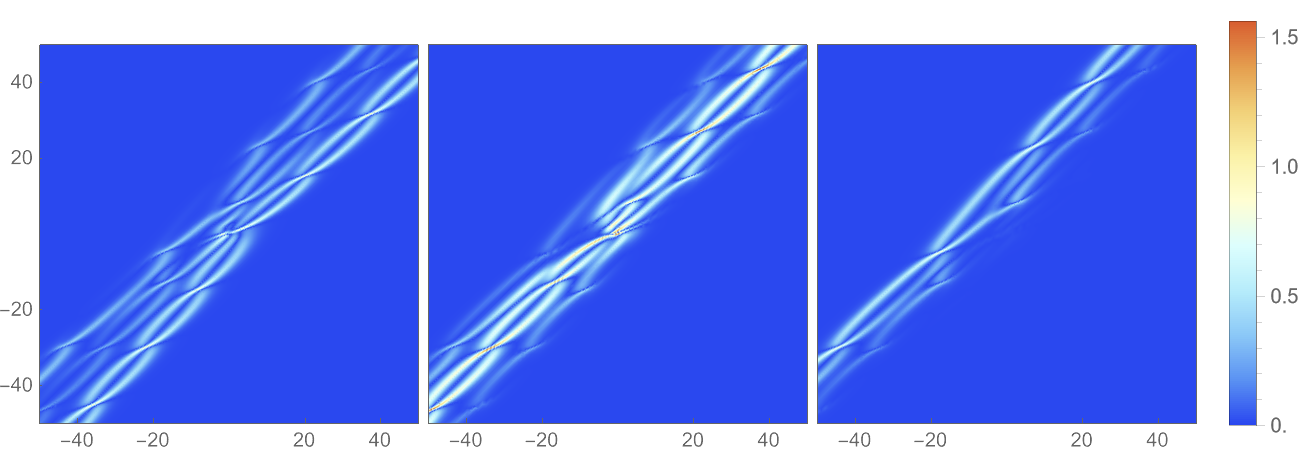}
\caption{ \label{QD31021}Density structures of three components $|q_1|$, $|q_2|$ and $ |q_{3}|$ (from left to right) as $N=1$, $\lambda_1=\frac{1}{20}+\frac{\mathrm{i}}{2}$, $m_1=2$, $\mathbf{f}(\lambda)=\left(1,\lambda-\lambda_1,(\lambda-\lambda_1)^2\right)^\mathrm{T}$.}
\end{figure}

For $N=1$, we have derived a suite of soliton solutions with varying complexity: a sixth-order pole  soltion solution is presented in Figure \ref{QD3115}, a third-order pole breather solution in Figure \ref{QD3102}, and a third-order pole chain-type soliton solution in Figure \ref{QD31021}.  The density structures of $(|q_1|,|q_2|,|q_3|)$ clearly demonstrate a pronounced dependence on the parameters we have meticulously selected. Specifically, varying configurations result in a different number of wave packets, emphasizing the pivotal role of parameter selection in shaping the soliton's composition, and also suggesting the subtle interplay and mutual influences among the components.

\begin{figure}[!htbp]
\centering
 \includegraphics[scale=.5]{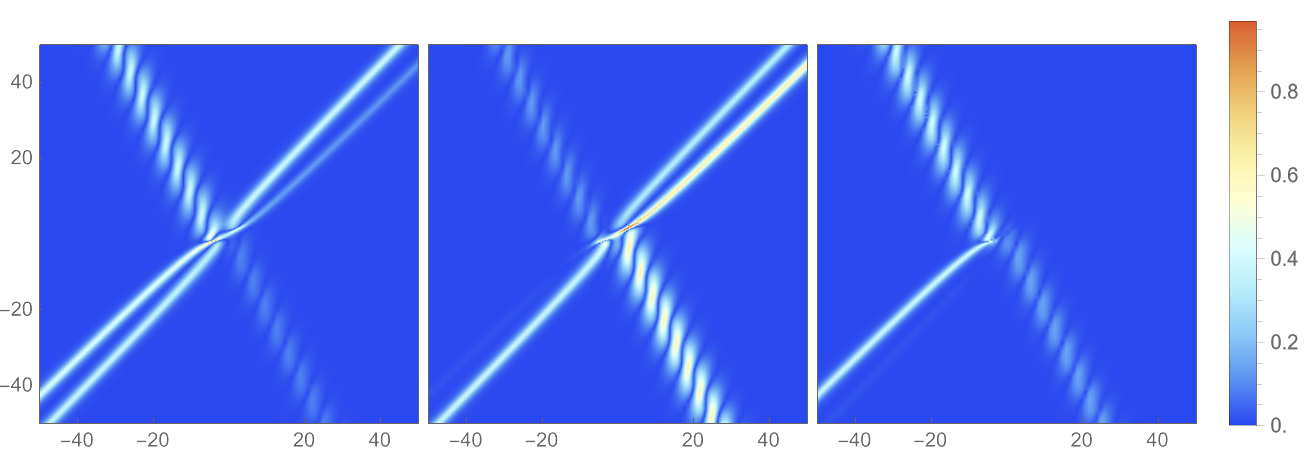}
\caption{ \label{QD32110} Density structures of three components $|q_1|$, $|q_2|$ and $ |q_{3}|$ (from left to right) as $N=2$, $\lambda_1=\frac{\mathrm{i}}{2}$, $\lambda_2=\frac{1}{4}+\frac{\mathrm{i}}{4}$, $m_1=1$, $m_2=0$, $\mathbf{f}(\lambda)=\left(\lambda^2,1,(\lambda-\lambda_1)^2\right)^\mathrm{T}$.}
\end{figure}

\begin{figure}[!htbp]
\centering
 \includegraphics[scale=.5]{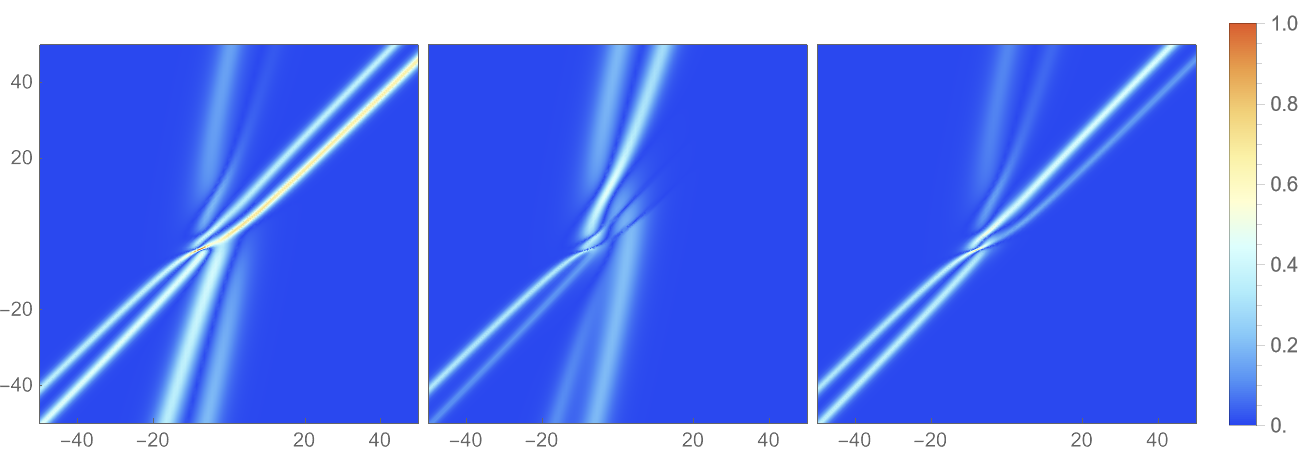}
\caption{ \label{QD32111}Density structures of three components $|q_1|$, $|q_2|$ and $ |q_{3}|$ (from left to right) as $N=2$, $\lambda_1=\frac{\mathrm{i}}{2}$, $\lambda_2=\frac{\mathrm{i}}{4}$, $m_1=1$, $m_2=1$, $\mathbf{f}(\lambda)=\left(\lambda^2,(\lambda-\lambda_1)^2,(\lambda-\lambda_2)^2\right)^\mathrm{T}$.}
\end{figure}

For $N=2$, we have derived a suite of soliton solutions with varying complexity: collision of a second-order pole soliton and a breather is presented in Figure \ref{QD32110}, collision of two second-order pole soliton solutions in Figure \ref{QD32111}.  The density structures of $(|q_1|,|q_2|,|q_3|)$ reveals distinct behaviors among the three components following the collision of two solitons: some retain their energy post-collision, others coalesce from two distinct wave packets into a single entity, and still others vanish completely after the interaction.
	\section*{Acknowledgment}
	This work was supported by National Natural Science Foundation of China (Grant Nos. 12171439 and 12101190).	
	
	\section*{Data availability}
	All data generated or analyzed during this study are including in this published article.	
	
	\section*{Declarations}	
	
	\section*{Conflict of Interest}		
	The authors declare that they have no conflict of interest.


\begin{thebibliography}{99}
   \bibitem{Newell1985}
    Newell  A.C.: Solitons in Mathematics and Physics.  SIAM, Philadelphia (1985)
   \bibitem{Hasegawa1995}
    Hasegawa  A., Kodama  Y.: Solitons in Optical Communications. Oxford University Press, New York (1995)
   \bibitem{Kuznetsov1986}
    Kuznetsov  E.A., Rubenchik  A.M., Zakharov  V.E.: Soliton stability in plasmas and hydrodynamics.  Phys. Rep. \textbf{142}, 103-165 (1986)
    \bibitem{Pathak2017}
    Pathak  P., Sharma  S.K., Nakamura  Y., Bailung  H.: Observation of ion acoustic multi-Peregrine solitons in
              multicomponent plasma with negative ions.  Phys. Lett. A \textbf{381}, 4011-4018 (2017)
    \bibitem{Cheemaa2020}
    Cheemaa  N., Seadawy  A.R., Sugati T.G., Baleanu  D.: Study of the dynamical nonlinear modified Korteweg--de Vries equation arising in plasma physics and its analytical wave solutions. Results Phys.\textbf{19}, 103480 (2020)
    \bibitem{Pitaevskii2003}
    Pitaevskii  L., Stringari  S.: Bose--Einstein Condensation. Oxford University Press, Oxford (2003)
    \bibitem{Osborne2010}
    Osborne  A.R.: Nonlinear Ocean Waves and the Inverse Scattering Transform. Elsevier, New York (2010)
     \bibitem{Gardner1967}
    Gardner  C.S., Greene  J.M., Kruskal  M.D., Miura  R.M.: Method for solving the Korteweg--de Vries equation. Phys. Rev. Lett. \textbf{19}, 1095-1097 (1967)
    \bibitem{Ablowitz1974}
    Ablowitz  M.J., Kaup  D.J., Newell  A.C., Segur  H.: The inverse scattering transform-Fourier analysis for
              nonlinear problems. Stud. Appl. Math. \textbf{53}, 249-315 (1974)
    \bibitem{Ablowitz1981}
    Ablowitz  M.J., Segur  H.: Solitons and the Inverse Scattering Transform. SIAM, Philadelphia (1981)
    \bibitem{Hirota1971}
    Hirota  R.: Exact solution of the Korteweg--de Vries equation for multiple   collisions of solitons. Phys. Rev. Lett. \textbf{27}, 1192-1194 (1971)
    \bibitem{Hirota2004}
    Hirota  R.: The Direct Method in Soliton Theory. Cambridge University Press, Cambridge (2004)
    \bibitem{Matveev1991}
    Matveev  V.B., Salle  M.A.: Darboux Transformations and Solitons. Springer, Berlin (1991)
    \bibitem{Gu2005}
    Gu  C.H., Hu  H.S., Zhou  Z.X.: Darboux Transformations in Integrable Systems: Theory and their Applications to Geometry. Springer, Dordrecht (2005)
     \bibitem{Guo2013}
    Guo  B.L., Ling  L.M., Liu  Q.P.: High-order solutions and generalized Darboux transformations
              of derivative nonlinear Schr\"odinger equations.  Stud. Appl. Math. \textbf{130}, 317-344 (2013)
    \bibitem{Novikov1984}
    Novikov  S., Manakov  S.V., Pitaevski\u i  L.P., Zakharov  V.E.: Theory of Solitons: The Inverse Scattering Method. Consultants Bureau Press, New York (1984)
   \bibitem{Its2003}
    Its  A.R.: The Riemann--Hilbert problem and integrable systems. Notices Amer. Math. Soc. \textbf{50}, 1389-1400 (2003)
    \bibitem{Zakharov1972}
    Zakharov  V.E., Shabat  A.B.: Exact theory of two-dimensional self-focusing and
              one-dimensional self-modulation of waves in nonlinear media. Soviet Phys. JETP \textbf{34}, 62-69 (1972)
    \bibitem{Olmedilla1987}
    Olmedilla  E.: Multiple pole solutions of the nonlinear Schr\"odinger equation. Phys. D \textbf{25}, 330-346 (1987)
    \bibitem{Aktosun2007}
    Aktosun  T., Demontis  F., van der Mee  C.: Exact solutions to the focusing nonlinear Schr\"odinger
              equation. Inverse Prob. \textbf{23}, 2171-2195 (2007)
    \bibitem{Schiebold2017}
    Schiebold  C.: Asymptotics for the multiple pole solutions of the nonlinear Schr\"odinger equation. Nonlinearity \textbf{30}, 2930-2981 (2017)
    \bibitem{Zhang2020a}
    Zhang  Y.S., Tao  X.X., Yao  T.T., He  J.S.: The regularity of the multiple higher-order poles solitons of
              the NLS equation. Stud. Appl. Math. \textbf{145}, 812-827 (2020)
    \bibitem{Bilman2019}
    Bilman  D., Buckingham  R.: Large-order asymptotics for multiple-pole solitons of the focusing nonlinear Schr\"odinger equation. J. Nonlinear Sci.\textbf{29} 2185-2229 (2019)
    \bibitem{Bilman2021}
    Bilman  D., Buckingham  R., Wang  D.S.: Far-field asymptotics for multiple-pole solitons in the
              large-order limit. J. Differ. Equ. \textbf{297}, 320-369 (2021)

    \bibitem{Wadati1982}
    Wadati  M., Ohkuma  K.: Multiple-pole solutions of the modified Korteweg--de Vries
              equation. J. Phys. Soc. Jpn. \textbf{51}, 2029-2035 (1982)
    \bibitem{Zhang2020b}
    Zhang  Y.S., Tao  X.X., Xu  S.W.: The bound-state soliton solutions of the complex modified
              KdV equation. Inverse Prob. \textbf{36}, 065003 (2020)
    \bibitem{Tsuru1984}
    Tsuru  H., Wadati  M.: The multiple pole solutions of the sine-Gordon equation. J. Phys. Soc. Jpn. \textbf{53}, 2908-2921 (1984)
    \bibitem{Fan2022}
    Chen  M.S., Fan  E.G.: Riemann--Hilbert approach for discrete sine-Gordon equation with simple and double poles. Stud. Appl. Math. \textbf{148}, 1180–1207  (2022)
    \bibitem{Liu2024a}
    Liu  H., Shen  J., Geng  X.G.: Riemann--Hilbert method to the Ablowitz--Ladik equation:
              higher-order cases. Stud. Appl. Math. \textbf{153}, e12748 (2024)
    \bibitem{Liu2024b}
    Liu  H., Zhou  P.P., Geng  X.G.: Inverse scattering transform for the Sasa--Satsuma
              equation: multiple-pole case of $N$ pairs. Z. Angew. Math. Phys. \textbf{75}, 193 (2024)
    \bibitem{Liu2024c}
    Liu  H., Shen  J., Geng  X.G.: Multiple higher-order pole solutions in spinor
              Bose--Einstein condensates. J. Nonlinear Sci. \textbf{34}, 48 (2024)
    \bibitem{Yao2004}
    Yao  R.X., Qu  C.Z., Li  Z.B.: Painlev\'e property and conservation laws of multi-component mKdV equations. Chaos Solitons Fractals \textbf{22}, 723-730 (2004)
    \bibitem{Zhang2008}
    Zhang  H.Q., Tian  B., Xu  T., Li  H., Zhang  C., Zhang  H.: Lax pair and Darboux transformation for multi-component modified Korteweg--de Vries equations. J. Phys. A \textbf{41}, 355210 (2008)
    \bibitem{Geng2014}
    Geng  X.G., Zhai  Y.Y., Dai  H.H.: Algebro-geometric solutions of the coupled modified Korteweg--de Vries hierarchy. Adv. Math. \textbf{263}, 123-153 (2014)
    \bibitem{Tian2017}
    Tian  S.F.: Initial-boundary value problems of the coupled modified Korteweg--de Vries equation on the half-line via the Fokas method. J. Phys. A \textbf{50}, 395204 (2017)
    \bibitem{Chang2018}
    Chang  X.K., He  Y., Hu  X.B., Li  S.H., Tam  H.W., Zhang  Y.N.: Coupled modified KdV equations, skew orthogonal polynomials, convergence acceleration algorithms and Laurent property. Sci. China Math. \textbf{61}, 1063-1078 (2018)
    \bibitem{Pelinovsky2018}
    Pelinovsky  D.E., Stepanyants  Y.A.: Helical solitons in vector modified Korteweg--de Vries equations. Phys. Lett. A \textbf{382}, 3165-3171 (2018)
    \bibitem{Geng2019}
    Geng  X.G., Chen  M.M., Wang  K.D.: Long-time asymptotics of the coupled modified Korteweg--de Vries equation. J. Geom. Phys. \textbf{142}, 151-167 (2019)
    \bibitem{Wurile2019}
    Wurile, Zhaqilao: Darboux transformation and soliton solutions for a three-component modified Korteweg--de Vries equation. Wave Motion \textbf{88}, 73-84 (2019)
    \bibitem{Adamopoulou2020}
    Adamopoulou  P., Papamikos  G.: Drinfel'd--Sokolov construction and exact solutions of vector modified KdV hierarchy. Nuclear Phys. B \textbf{952}, 114933 (2020)
    \bibitem{Xiao2021}
    Xiao  Y., Fan  E.G., Liu  P.: Inverse scattering transform for the coupled modified Korteweg--de Vries equation with nonzero boundary conditions. J. Math. Anal. Appl. \textbf{504}, 125567 (2021)
    \bibitem{Malham2022}
    Malham  S.J.A.: The non-commutative Korteweg--de Vries hierarchy and combinatorial Pöppe algebra. Phys. D \textbf{434}, 133228 (2022)
    \bibitem{Liu2023a}
    Liu  N., Zhao  X.D., Guo  B.L.: Long-time asymptotic behavior for the matrix modified Korteweg--de Vries equation. Phys. D \textbf{443}, 133560 (2023)
    \bibitem{Ye2023}
    Ye  R.S., Zhang  Y.: Initial-boundary value problems for the two-component complex modified Korteweg--de Vries equation on the interval. Discrete Contin. Dyn. Syst. Ser. S \textbf{16}, 671-707 (2023)
    \bibitem{Wong2024}
    Wong  C.N., Yin  H.M., Chow  K.W.: Transient modes for the coupled modified Korteweg--de Vries equations with negative cubic nonlinearity: stability and applications of breathers. Chaos \textbf{34}, 083132 (2024)
    \bibitem{Liu2016}
    Liu  H., Geng  X.G.: Initial-boundary problems for the vector modified Korteweg--de Vries equation via Fokas unified transform
              method. J. Math. Anal. Appl. \textbf{440}, 578-596 (2016)
    \bibitem{Wang2020}
    Wang  X.B., Han  B.: Application of the Riemann--Hilbert method to the vector modified
   Korteweg--de Vries equation. Nonlinear Dyn. \textbf{99}, 1363-1377 (2020)
    \bibitem{Liu2023b}
    Liu  H., Shen  J., Geng  X.G.: Inverse scattering transformation for the $N$-component
              focusing nonlinear Schr\"odinger equation with nonzero
              boundary conditions. Lett. Math. Phys. \textbf{113}, 23 (2023)
     \bibitem{Zhou1989}
    Zhou  X.: The Riemann--Hilbert problem and inverse scattering. SIAM J. Math. Anal. \textbf{20}, 966-986 (1989)
     \bibitem{Fokas2008}
    Fokas  A.S.: A unified approach to boundary value problems. SIAM, Philadelphia (2008)


\end{thebibliography}
\end{document}